\documentclass[journal,twoside,10pt,final,twocolumn,a4paper]{IEEEtran}

\usepackage[pagebackref=false,breaklinks=true,colorlinks,bookmarks=false]{hyperref}
\hypersetup
{
    pdfauthor={Vikram Singh and Anurag Mittal},
    pdfsubject={Arxiv Preprint},
    pdftitle={WDN: A Wide and Deep Network to Divide-and-Conquer Image Super-resolution},
    pdfkeywords={Image super-resolution, Image restoration, Wide and deep network, WDN, Divide-and-conquer}
}

\usepackage[pdftex]{graphicx}
\usepackage{framed,multirow}
\usepackage{amssymb}
\usepackage{latexsym}
\usepackage{url}
\usepackage{xcolor}
\usepackage{subcaption}
\usepackage[stable]{footmisc}
\usepackage{float}
\usepackage{soul}
\usepackage{relsize}
\usepackage{mathtools}
\usepackage{cite}
\newcommand{\etal}{\MakeLowercase{\textit{et al.}}}

\begin{document}

\title{WDN: A Wide and Deep Network to Divide-and-Conquer Image Super-resolution}

\author{Vikram~Singh,~\IEEEmembership{Graduate Student Member,~IEEE,}
Anurag~Mittal,~\IEEEmembership{Senior Member,~IEEE}% 
\thanks{Both the authors are from Computer Vision Lab, Department of Computer Science and Engineering, Indian Institute of Technology, Madras,
Chennai, 600036 India. The corresponding author is Vikram Singh, e-mail: vsingh@cse.iitm.ac.in.}% 
}

\markboth{Singh and Mittal : WDN: A wide and deep network to divide-and-conquer image super-resolution}%
{Singh and Mittal : WDN: A wide and deep network to divide-and-conquer image super-resolution}

\maketitle

\begin{abstract}
Divide and conquer is an established algorithm design paradigm that has proven itself to solve a variety of problems efficiently. However, it is yet to be fully explored in solving problems with a neural network, particularly the problem of image super-resolution. In this work, we propose an approach to divide the problem of image super-resolution into multiple sub-problems and then solve/conquer them with the help of a neural network. Unlike a typical deep neural network, we design an alternate network architecture that is much wider (along with being deeper) than existing networks and is specially designed to implement the divide-and-conquer design paradigm with a neural network. Additionally, a technique to calibrate the intensities of feature map pixels is being introduced. Extensive experimentation on five datasets reveals that our approach towards the problem and the proposed architecture generate better and sharper results than current state-of-the-art methods.
\end{abstract}


\IEEEpeerreviewmaketitle

\section{Introduction}
\IEEEPARstart{I}{mage} super-resolution is one of the challenging restoration tasks that involves increasing the resolution of the given image. Recent technological advances in the domain of display devices (e.g. high/ultra-high-definition screens) and enormous availability of low-resolution images (captured by old cameras/mobile-phones) have made this problem to garner significant research attention from the Computer Vision community. Virtual Super Resolution (VSR) technology from AMD, Dynamic Super-Resolution (DSR) technology developed by Nvidia and the most recent Nvidia's Deep Learning Super Sampling (DLSS 2.0) technology are a few examples that highlight the commercial importance and viability of the super-resolution techniques.

Substantial progress has already been accomplished in solving the image super-resolution problem. Notably, the existing techniques follow the general principle of `building deeper networks and training them on large data'. For instance, the residual channel attention network proposed by Zhang \etal \cite{10.1007/978-3-030-01234-2_18}, multi-scale residual architecture proposed by Li \etal \cite{Li_2018_ECCV}, and second-order attention network proposed by Dai \etal \cite{Dai_2019_CVPR}, among others still use this approach. However, we observe that such techniques still have a significant scope left for improvement, specifically in terms of improving the quality of the upsampled results, that primarily lack the required sharpness.

In this work, we attempt to improve upon the performance of existing image super-resolution methods with the motivation that the well-established approach of divide-and-conquer when applied with a neural network for image super-resolution might provide a performance gain. With this approach, we divide the image super-resolution problem into multiple sub-problems and solve them individually, thereby merging the sub-solutions to generate the final solution/upsampled-image. Unlike existing image super-resolution networks that are deep, we build an alternate network architecture that is specifically designed to work on the `divide and conquer' design paradigm and hence, is much wider along with being deeper.

A wide neural network that is designed to divide and conquer the problem has the advantage of better learning and faster processing. It can either execute on a single powerful GPU or multiple smaller GPUs in parallel. As such, we can divide the given complex problem into many simpler sub-problems, and then multiple sub-networks that are connected along the wide network's width can be trained simultaneously to solve those simpler sub-problems. This training will increase the expertise of the sub-networks towards solving sub-problem of a particular type and will ultimately improve the overall network performance. We elaborate more on this point in Section \ref{sec:wdn} after describing the functioning and architecture of our network. Nevertheless, as our proposed network architecture has much more width than existing networks, we name it \textit{WDN} to represent its wide and deep design. 

An arduous challenge that is encountered while solving the image super-resolution problem is of predicting/upsampling the image while maintaining its `sharpness'. The generation of a sharp prediction requires precise prediction of the high-frequency details in the image. These details are found in the regions of the high spatial gradient, for instance, the object edges. High-frequency prediction is challenging as these are the details that suffer most of the losses in a low-resolution image as compared to low-frequency details. Unless explicit measures are not enforced to predict high-frequency details, there is a high probability that the predicted image will lack the required sharpness. Indeed, this is observed in most of the current methods that do not enforce these measures explicitly. To overcome this challenge and to generate sharper results, we upsample the image in two parts by `separately predicting its high-frequency and low-frequency channels'. The separately predicted high-frequency and low-frequency channels are then fused to generate the final output.

Apart from a wider design (based on divide-and-conquer) and explicit measure for predicting high-frequency details, we also introduce a method to calibrate the feature-map pixels that are generated inside the network. This method calibrates the intensity of pixels using a self-learned pixel relevance value to improve the overall network performance. We further elaborate on this in Section \ref{subsec:wdn_arch}.

Before proceeding ahead, we mention that this work is an expansion of our prior work\footnote{Singh \etal, "Going Much Wider with Deep Networks for Image Super-Resolution," IEEE/CVF WACV, 2020, pp. 2332-2343, DOI: \href{https://dx.doi.org/10.1109/WACV45572.2020.9093317}{10.1109/WACV45572.2020.9093317}.} The unique contributions of this article over the conference version are: 1) Division of problem based on scale factor (i.e. two 2$\times$ upsampling), 2) Use of attention mechanism to better combine sub-solutions, 3) Weight sharing blocks to reduce the number of parameters, 4) Improved architecture that requires minimisation of only 12 losses as opposed to 49 losses of the network proposed in the conference version, 5) Intermediate layer output visualisations for better understanding of the network functioning, 6) Extensive literature survey to cover a large set of existing works, 7) Additional experiment on Manga109 dataset, 8)  Analysis of model complexity and execution time, 9) Discussion on the choice of high-frequency extractor: Sobel, 10) Analysis of the DAC based design, and 11) Evaluation of different training procedures. We refer the reader to the conference version and to the rest of this article for a better understanding of the stated contributions. Next, we discuss some state-of-the-art methods for image super-resolution.

\section{Prior Work}
\label{sec:related_work}
Methods described in \cite{5459271,5466111,6751349,7299003} comprise the earlier approach for image super-resolution that are mostly non-deep learned. Glasner \etal \cite{5459271} perform upsampling by using internal recurrence of image patches. Yang \etal \cite{5466111} consider image patches as sparse signal representations. Timofte \etal \cite{6751349} reduce the execution time of existing super-resolution techniques by using sparse dictionaries and neighbourhood regression, while Schulter \etal \cite{7299003} upsample by coining a random forest-based approach.

Ideas proposed in \cite{7780576,10.1007/978-3-319-46475-6_43,7780551,10.1007/978-3-319-46454-1_20,7780550} attempt to establish a non-linear connection between the high-resolution output and the corresponding low-resolution input using deep neural networks. Though the performance of these works is decent, they also have a large number of trainable parameters, and they lack in predicting fine textures in images. Tai \etal \cite{8099781} reduce the parameters by recursive modelling, whereas \cite{8100101,10.1007/978-3-030-01234-2_18,8578422,10.1007/978-3-030-01249-6_16} adopt residual learning to predict fine textures. Adversarially trained networks (GANs) have also been deployed to increase the sharpness of the prediction in \cite{8237743,8099502,8237298}. These techniques indirectly highlight the importance of predicting the high-frequency details as without them the results become even more blurred. However, none of these techniques take explicit measures for their prediction to make the results sharper. High-frequency details are responsible to bring-in sharpness as they comprise the fine-textures of the predicted image. We have designed WDN to predict the high-frequency details explicitly, and thus the results that it generates are visually sharper than the current state-of-the-art.

To adapt to high magnification factors, Dahl \etal \cite{8237843} propose a deep probabilistic network. Tai \etal \cite{8237748} present a memory block architecture to solve the long dependency problem (i.e. the influence of initial layers on the final prediction) of deep networks. The work of Lim \etal \cite{8014885} aims to reduce the modules in conventional residual networks. Zheng \etal \cite{10.1007/978-3-030-01231-1_6} attempt to optimise the performance of the model using pixel-level alignment. Hui \etal \cite{8578180} give a deep yet compact network that directly predicts high-resolution images from the low-resolution input. Li \etal \cite{10.1007/978-3-030-01237-3_32} have convolutional kernels of multiple sizes to identify the image features on different scales dynamically. Using wavelets, the model by Zhong \etal \cite{Zhong:2018:JSL:3326943.3326959} predict the high-resolution image with better textural details. Though these methods perform sufficiently well with the Bi-cubically downsampled low-resolution images, their performance deteriorates when the input comes from a real low-resolution camera. Bulat \etal \cite{10.1007/978-3-030-01231-1_12} attempt to address this with GANs, whereas Zhang \etal \cite{8578442} design a model for input with multiple and spatially variant degradations. Our divide-and-conquer based approach of WDN towards the super-resolution problem significantly differs from all the cited works. WDN has been tested to work with the most commonly used Bi-cubic downsampling for a fair comparison with the state-of-the-art.

Some authors have built divide-and-conquer networks to solve problems such as exclusive-or, clustering, manipulation, locomotion, and super-resolution, among others. An earlier (1993) network for exclusive-or by Romaniuk and Hall \cite{ROMANIUK19931105} trains at the cellular level and unlike WDN, it has no back-propagation or weight sharing. Nowak \etal \cite{nowak2018divide} give a bi-modular network for clustering. Unlike multi-modular WDN, it recursively divides the problem to build a binary tree of sub-problems. The approach of Ghosh \etal \cite{ghosh2018divideandconquer} for tasks of manipulation and locomotion use divide-and-conquer with reinforcement learning. Significantly different, WDN is a divide-and-conquer based wide and deep learned network.

Lin \etal \cite{9010947}, Kim \etal \cite{AAAI2020}, and Huang \etal \cite{huang2019divideandconquer} propose divide-and-conquer based adversarially trained networks for generation, super-resolution, and enhancement of images, respectively. The first generates an image by concatenating spatially-separated predicted patches. Second, divide the upsampling problem into only three sub-problems of reconstruction, detail restoration, and local contrast enhancement. They train their network in a unified manner. Third, divides the problem at three levels of perception, frequency and dimension. As opposed to these works, the sub-solutions predicted by WDN are not localised patches but they contain pixels that are evenly distributed across the entire spatial space of the model-prediction. WDN statically divides the 4$\times$ upsampling problem (i.e. ground-truth) into 11 sub-problems based on scale and frequency and the model input into 32 sub-inputs. The frequency division of WDN is also different and is performed with Sobel filters. Each level of WDN follows a similar division pattern based on scale and frequency. WDN does not take explicit measures for contrast-enhancement. We show later that with these difference, WDN is able to generate better and sharper results that state-of-the-art. The design of WDN facilitates faster processing on a multi-GPU system.

We now discuss some of the recent methods image super-resolution methods and compare WDN with them all-together at the end. The following techniques focus on varied aspects of the super-resolution problem. For instance, the findings of Han \etal \cite{8578276} reveal that a single-state recurrent neural network (RNN) could approximate many deep super-resolution networks. The authors of \cite{Gu_2019_CVPR,Zhou_2019_ICCV1,Zhang_2019_CVPR1} attempt to estimate the unknown blur kernel for blind super-resolution. Qiu \etal \cite{Qiu_2019_ICCV} work on texture super-resolution. He \etal \cite{He_2019_CVPR} devises an ODE-inspired scheme. Haris \etal \cite{8578277} come up with an iterative up-down sampling mechanism. Hu \etal \cite{Hu_2019_CVPR} introduce a flexible model that can accept a scale factor along with the input to compute the network parameters according to the scale factor dynamically. Li \etal \cite{Li_2019_CVPR} refine the low-level representations with high-level information, Dai \etal \cite{Dai_2019_CVPR} propose a model that captures the long-distance dependencies along with structural information by embedding non-local operations in the network to account for the correlation between features of different layers. Significantly different from these, Shocher \etal \cite{8578427} design an unsupervised approach to train the network at test time with the test image itself. 

Park \etal \cite{10.1007/978-3-030-01270-0_27} suggest that deploying a GAN to predict fine textures also amplifies the high-frequency noise. They address this issue by using an additional discriminator that keeps a check on the amplified noise. Zhang \etal \cite{Zhang_2019_ICCV} along with Rad \etal \cite{Rad_2019_ICCV} also deploy GANs that focus on the generator and discriminator respectively. Some works are also focused on reducing the model size for instance: the lightweight network of Liu \etal \cite{9053890} with progressive residual learning, residual global context network of Liu \etal \cite{9054003} that achieves a better trade-off between the number of parameters \& the upsampling quality, and the discriminant information pruning criteria based network of Hou \etal \cite{9054019}.

The most recent works on image super-resolution includes the multi-path adaptive modulation network of Kim \etal \cite{KIM2020} that modulate the residual feature responses, Channel splitting and fusion network of Zou \etal \cite{9053516} that obtain the respective contribution of each channel for predicting the result.  With a different focus from these, Wang \etal \cite{9066958} analyse multiple Gaussian degradations in an attempt to reduce the reconstruction error in real-world data, Qin \etal \cite{Qin2020} combine the ideas of the channel, and spatial attention for building a deep multilevel residual attention network and lastly, Wu \etal \cite{Wu2020} come up with a novel perceptual loss for upsampling. Though most recent, however, none of these techniques were able to cross the benchmark established by an earlier but current state-of-the-art method \cite{Qiu_2019_ICCV}.

To the best of our knowledge, WDN predominantly differs from the above-cited methods in its approach towards solving the image super-resolution problem. Notably, the significant differences/contribution of our work are:
\begin{itemize}
    \item Our work adopts an established algorithm design paradigm `divide-and-conquer' to solve the super-resolution problem by dividing it into multiple sub-problems.
    \item Our work proposes a much wider and deep network architecture that can solve the sub-problems separately and parallelly on one or more GPUs.
    \item Our work introduces a technique to calibrate the intensities of pixels in feature maps that subsequently improve the upsampling results with deeper networks.
\end{itemize}

\section{Wide and Deep Network (WDN)}
\label{sec:wdn}
In this paper, we propose the design of a wide and deep network (WDN) that solves the problem of image super-resolution by implementing the well-established paradigm of `divide-and-conquer'. Following this approach, our search for a better solution begins with the division of a given single problem into multiple sub-problems. The division that we make to create sub-problems is primarily based on the: 1) Frequency of the data and 2) Scaling factor.

In terms of frequency, the problem of predicting an upsampled image is divided into two sub-problems of separately predicting the high-frequency and low-frequency channels. A network that predicts high-frequency and low-frequency details together may become biased and drift towards predicting low-frequency details with more accuracy at the cost of a lower accuracy in predicting the high-frequency details resulting in blurry predictions. This is because typically an image has much more low-frequency details than high-frequency details. By dividing the problem explicitly into two separate problems of predicting high and low frequencies, the network can learn to gain better expertise (conquer) in high-frequency prediction (without low-frequency bias) subsequently generating sharper predictions.

In terms of the scaling factor, the 4$\times$ upsampling problem is divided into two successive sub-problems of 2$\times$ upsampling. The scaling factor division divides the problem into multiple sub-problems wherein multiple sub-networks can be deployed to gain better expertise in solving a particular sub-problem and generate better sub-solutions. Better sub-solutions can generate better solutions when combined. Moreover, such a division eases the training process by making the sub-networks executable on multiple GPUs in parallel, facilitating faster processing even with heavier sub-networks. 

With the above motivations for using the divide-and-conquer design paradigm, we combine the two criteria and divide the problem of 4$\times$ upsampling into three sets of eleven sub-problems as described in the next section.

\begin{figure*}[t]
\centering
 \includegraphics[width=0.95\linewidth]{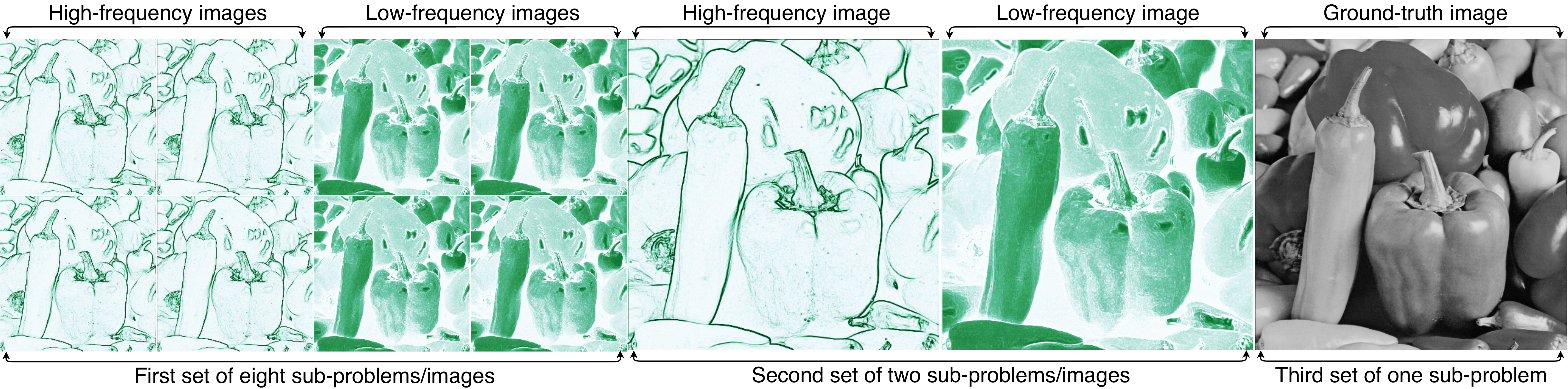}
\caption{Visualisation of the three sets of 11 sub-problems for 4$\times$ upsampling. The first set consists of the images at a scale of two, the second and the third set comprises of the images at a scale of four. Section \ref{subsec:subprb} describes the procedure to generate these sub-problems from the given ground-truth.}
\label{fig:sub_prb}
\end{figure*}

\subsection{Eleven sub-problems of 4$\times$ upsampling}
\label{subsec:subprb}

\begin{figure}
\centering
\begin{subfigure}{0.37\textwidth}
 \centering
 \includegraphics[width=0.99\linewidth]{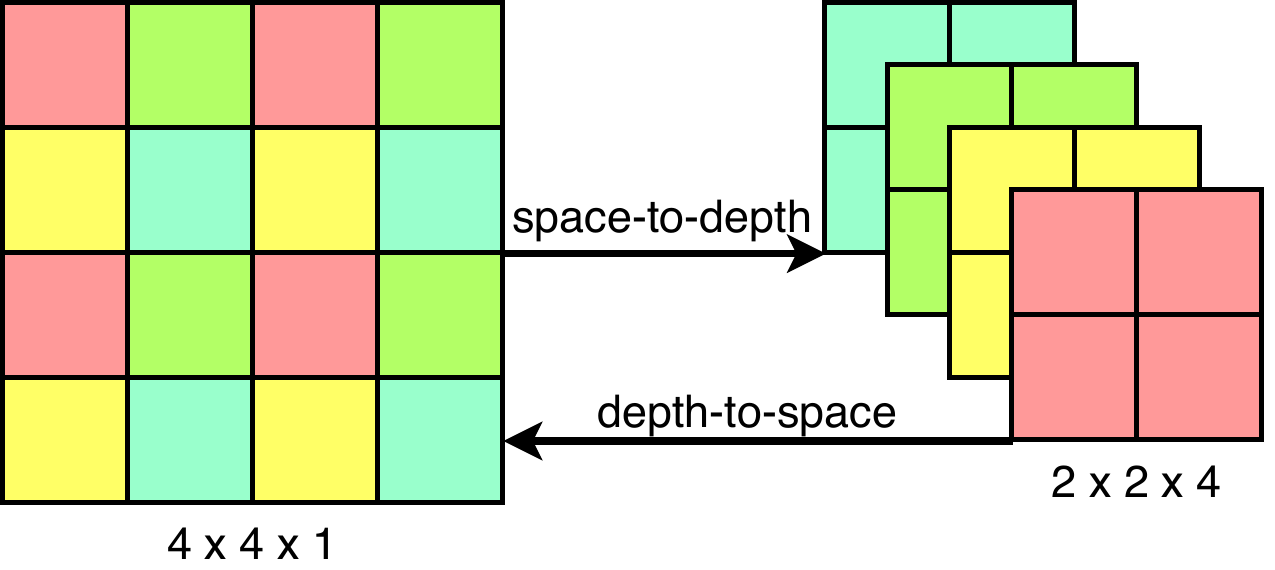}
 \caption{Block size of two.}
 \label{fig:s2d2}
\end{subfigure}

\vspace{0.5cm}

\begin{subfigure}{0.37\textwidth}
 \centering
 \includegraphics[width=0.99\linewidth]{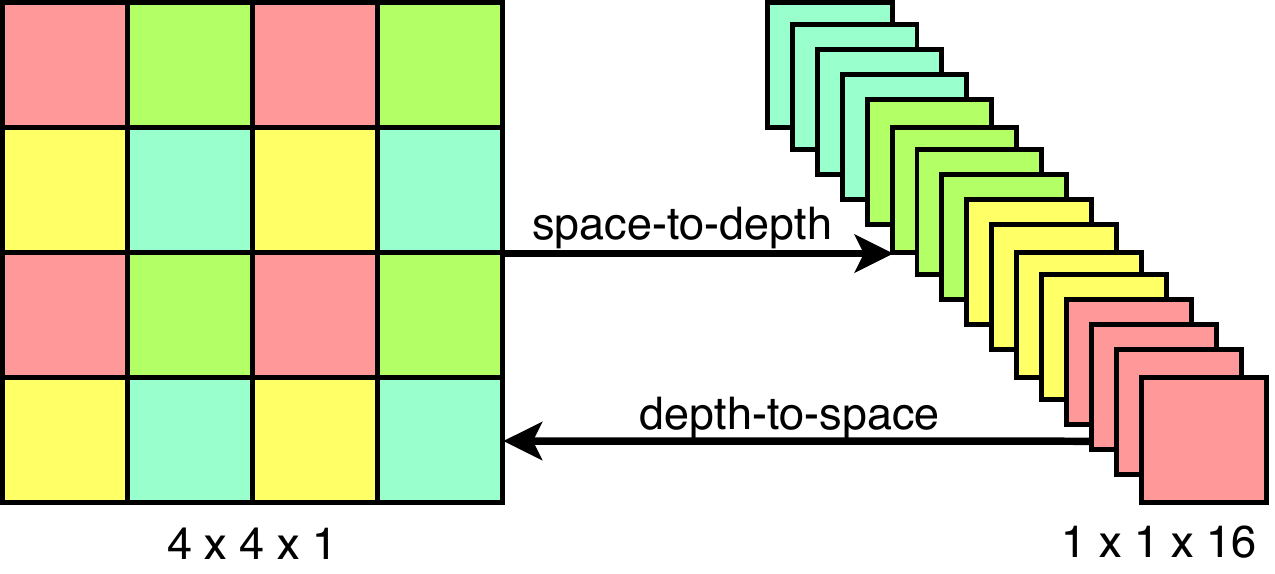}
 \caption{Block size of four.}
 \label{fig:s2d4}
\end{subfigure}
\caption{Illustration of space-to-depth and depth-to-space operations with different block sizes. \textbf{This figure is understandable in colour only.}}
\label{fig:s2d}
\end{figure}

A 4$\times$ upsampling problem consists of predicting a single image with 4$\times$ the resolution of the given image. In supervised paradigm, the ground-truth image with 4$\times$ resolution is available and the problem is to make the model prediction as close as possible to it in terms of some performance metric, e.g., PSNR \cite{IRANI1993324} and SSIM \cite{1284395}. Typically, neural networks make this prediction directly. However, instead of solving the problem of predicting the 4$\times$ image directly, we divide the problem into three sets of 11 sub-problems, as shown in Fig. \ref{fig:sub_prb} and train our network WDN to solve/predict them set-by-set. The first set of sub-problems consists of predicting four high-frequency and four low-frequency channels at a scale of two. To generate the ground-truths for the first set, Sobel filter is applied on the given 4$\times$ model ground-truth by following the procedure described in Section \ref{subsec:ehfm}. The procedure generates two 4$\times$ channels that contain the high-frequency and low-frequency details of the ground-truth image separately. Next, Space-to-depth \cite{space-to-depth} operator (illustrated in Fig. \ref{fig:s2d2}) is applied on both the generated high-frequency and low-frequency channels. In return, this operators gives four images corresponding to each of the 4$\times$ high-frequency and 4$\times$ low-frequency channels but on a scale factor of two. Together these four high-frequency and four low-frequency channels at a scale of two comprise the first set of sub-problems that are to be initially predicted to solve the problem of 4$\times$ upsampling later.

The second set of sub-problems include prediction of the high-frequency and low-frequency channels at a scale of four. These channels have already been generated/extracted from the given ground-truth while generating the sub-problems of the first set (before application of space-to-depth), and lastly, the third set contains a single problem of predicting the given 4$\times$ model ground-truth itself. Succinctly, the 4$\times$ upsampling problem gets divided into three sets of 11 (8 + 2 + 1) sub-problems that WDN is required to predict. Among these 11 sub-problems, five are exclusively for predicting the high-frequency details that explicitly helps to generate a sharper prediction, five are for predicting only the low-frequency details, and the last sub-problem is of predicting the desired outcome. Our empirical observations presented later in Section \ref{sec:experiment} reveal that this approach of subdividing the problem into multiple sub-problems is constructive and help in generating better and sharper results than the current state-of-the-art methods.

\subsubsection{Procedure to separate the frequency channels}
\label{subsec:ehfm}
Frequency channels are separated from an image `I' (scaled between 0-1), by application of Sobel filters $M$ and $N$. The application generates $\mathbb{D}_m$ \& $\mathbb{D}_n$, that represent the derivative approximations for horizontal and vertical changes respectively.
\begin{align*}
M = \begin{bmatrix} 
 +1 & 0 & -1 \\
+2 & 0 & -2 \\
+1 & 0 & -1 
\end{bmatrix}, \quad
N = \begin{bmatrix} 
 +1 & +2 & +1\\
 0 & 0 & 0 \\
-1 & -2 & -1
\end{bmatrix}
\end{align*}
Next, $\mathbb{D}_m$ \& $\mathbb{D}_n$ are used to obtain the high-frequency and low-frequency channels:
\begin{equation}
\label{eqn:sobel}
\centering
\begin{split}
\text{High-frequency Channel} = \mathbb{S}\:\bigg(\sqrt{ {\mathbb{D}^2_m} + {\mathbb{D}^2_n}}\bigg)\\
\text{Low-frequency Channel = I - High-frequency Channel}\\
\text{$\mathbb{S}$ scales the values in the range $[0,1]$}
\end{split}
\end{equation}
Extraction of low-frequency channel by subtracting Sobel extracted high-frequency channel from image `I' guarantees that the sum of low-frequency and high-frequency channels will give back the image `I'. We further discuss our choice of the channel extraction filter in Section \ref{subsec:sobel}. With an understanding of the decomposition of the 4$\times$ upsampling problem into multiple sub-problems, we now elaborate the functioning of WDN to solve these sub-problems.

\begin{figure*}[t]
\centering
 \includegraphics[width=0.95\linewidth]{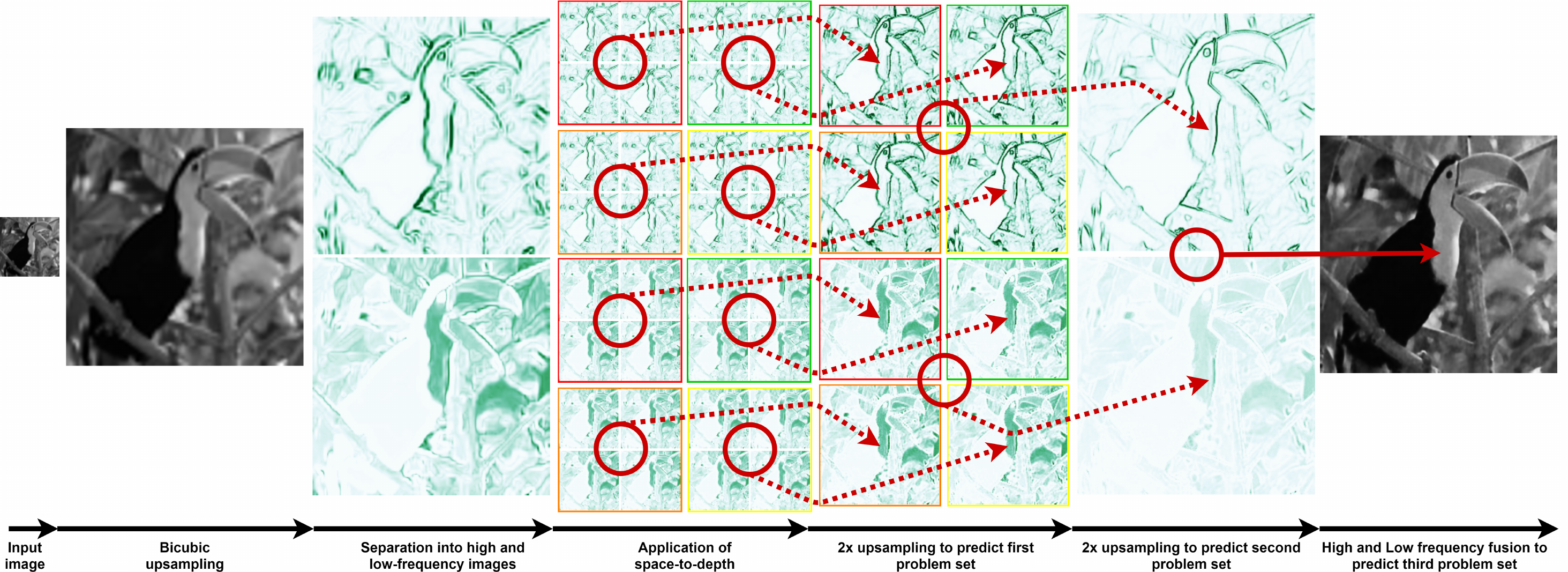}
\caption{Illustration of the functioning of WDN. The illustration is explained in Section \ref{subsec:functioning}. The high-frequency and low-frequency channels are in green-scale rather than grey-scale for better visualisation. This figure is better viewed on-screen after zooming.}
\label{fig:ga}
\end{figure*}

\subsection{Functioning of WDN}
\label{subsec:functioning}
WDN has been designed to accept and upsample (4$\times$) only the Luminance (Y) channel in the YCbCr colour-space of the image as human beings have high sensitivity towards a change in the Luminance. Similar to the work of Shi \etal \cite{7780576}, Kappeler \etal \cite{7444187}, and Liu \etal \cite{8237536}, the remaining channels are upsampled using a simple bi-cubic interpolation.

The functioning of WDN has been illustrated in Fig. \ref{fig:ga}. It can be seen that WDN starts functioning with a low-resolution input image. The first step consists of upsampling (4$\times$) the input using bi-cubic interpolation. In the second step, WDN separates (Ref. Section \ref{subsec:ehfm}) the upsampled image into two different channels having high-frequency and low-frequency detail, respectively. In the third step, pixels of the separated channels are further divided into 16 low-resolution channels, each by applying space-to-depth operator with a block size of four as illustrated in Fig. \ref{fig:s2d4}. The fourth step consists of consuming the channels in groups of four for predicting the solution of the first set of sub-problems, i.e. prediction of four 2$\times$ upsampled high-frequency channels and four 2$\times$ upsampled low-frequency channels. The fifth step is similar to the fourth step in processing. WDN once again consumes the 2$\times$ upsampled channels for predicting the solution of the second set of sub-problems, i.e. prediction of one 4$\times$ upsampled high-frequency channel and one 4$\times$ upsampled low-frequency channel. Lastly, in the sixth step, WDN consumes the two 4$\times$ upsampled high-frequency and low-frequency channels for fusing and predicting the solution of the third set of sub-problem, i.e. prediction of a single 4$\times$ image or the final model prediction. We now proceed to describe the architecture of WDN that has been designed to implement the described functionality.

\subsection{Architecture of WDN}
\label{subsec:wdn_arch}
\begin{figure*}[t]
\centering
\includegraphics[width=0.95\linewidth]{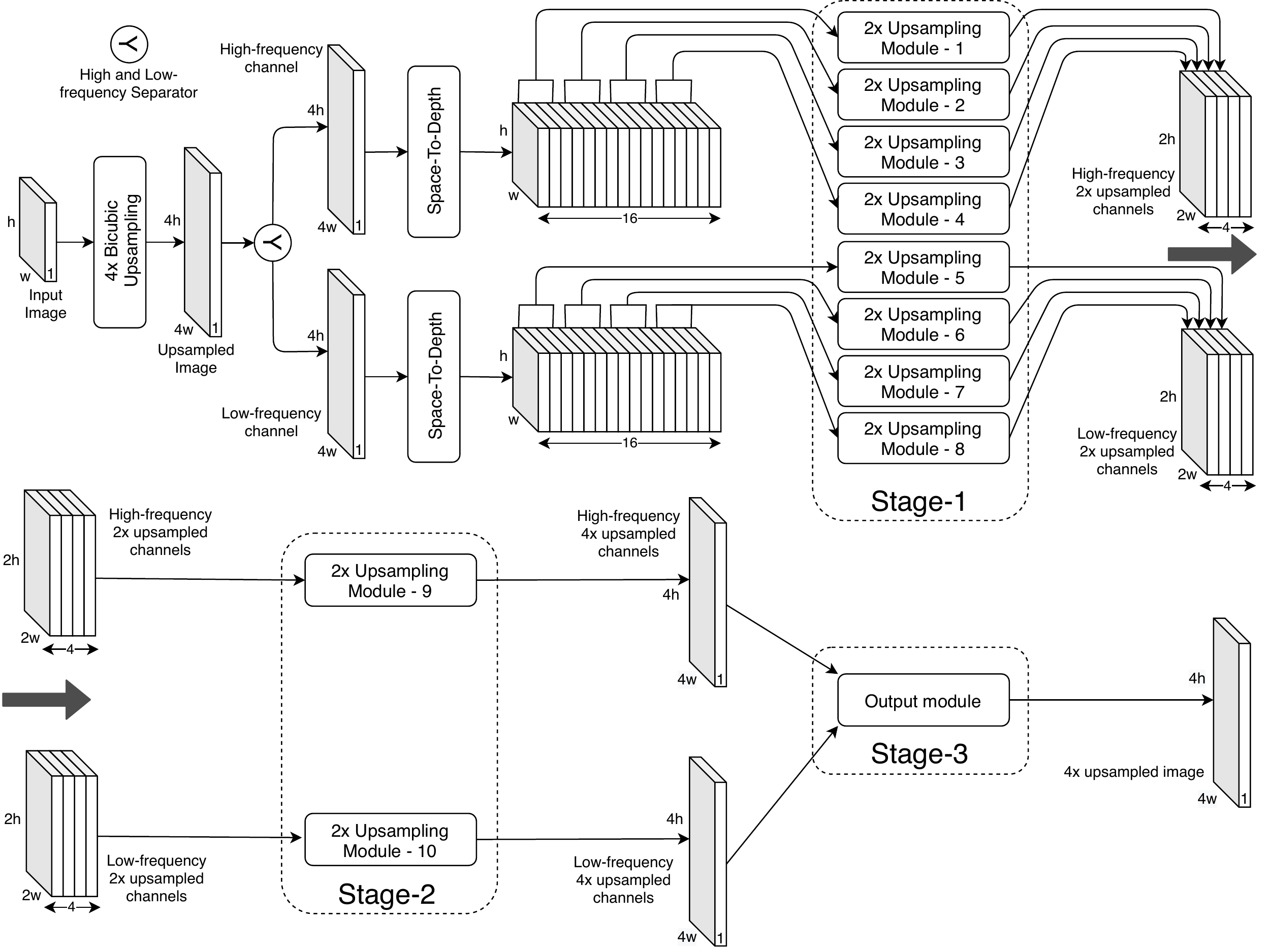}
\caption{Illustration showing the connectivity of the three stages in wide and deep architecture of the proposed super-resolution network WDN. The architecture is described in Section \ref{subsec:wdn_arch} and the detailed architecture of a 2$\times$ upsampling module is visualised in Fig. \ref{fig:network}}
\label{fig:model}
\end{figure*}

WDN has a wide and deep architecture with two non-trainable operations followed by three trainable stages that are connected in sequence as visualised in Fig. \ref{fig:model}. These stages have been designed to conquer/solve the sub-problems formulated in Section \ref{subsec:subprb} and implement the functionality described in the previous section. The non-trainable operations consist of the bicubic upsampling and the channel separation procedure. The first two stages perform a 2$\times$ upsampling on their respective inputs to eventually perform the required 4$\times$ upsampling while the third stage generates the desired output from the network.

In more detail, the first stage has a set of eight parallel 2$\times$ upsampling modules that each accepts and processes (in parallel) four channels to generate a single channel of double the input size. This design makes the first stage to have a width built with 32 (8$\times$4) deep networks connected in parallel that processes 32 input channels (16 of high-frequency and 16 of low-frequency) to generate eight output channels (four of high-frequency and four of low-frequency). Similarly, the second stage consists of two 2$\times$ upsampling modules that together accept eight channels generated by Stage-1 to generate two channels (one of high-frequency and one of low-frequency) of double the input size.

\begin{figure}[t]
\centering
 \includegraphics[width=0.95\linewidth]{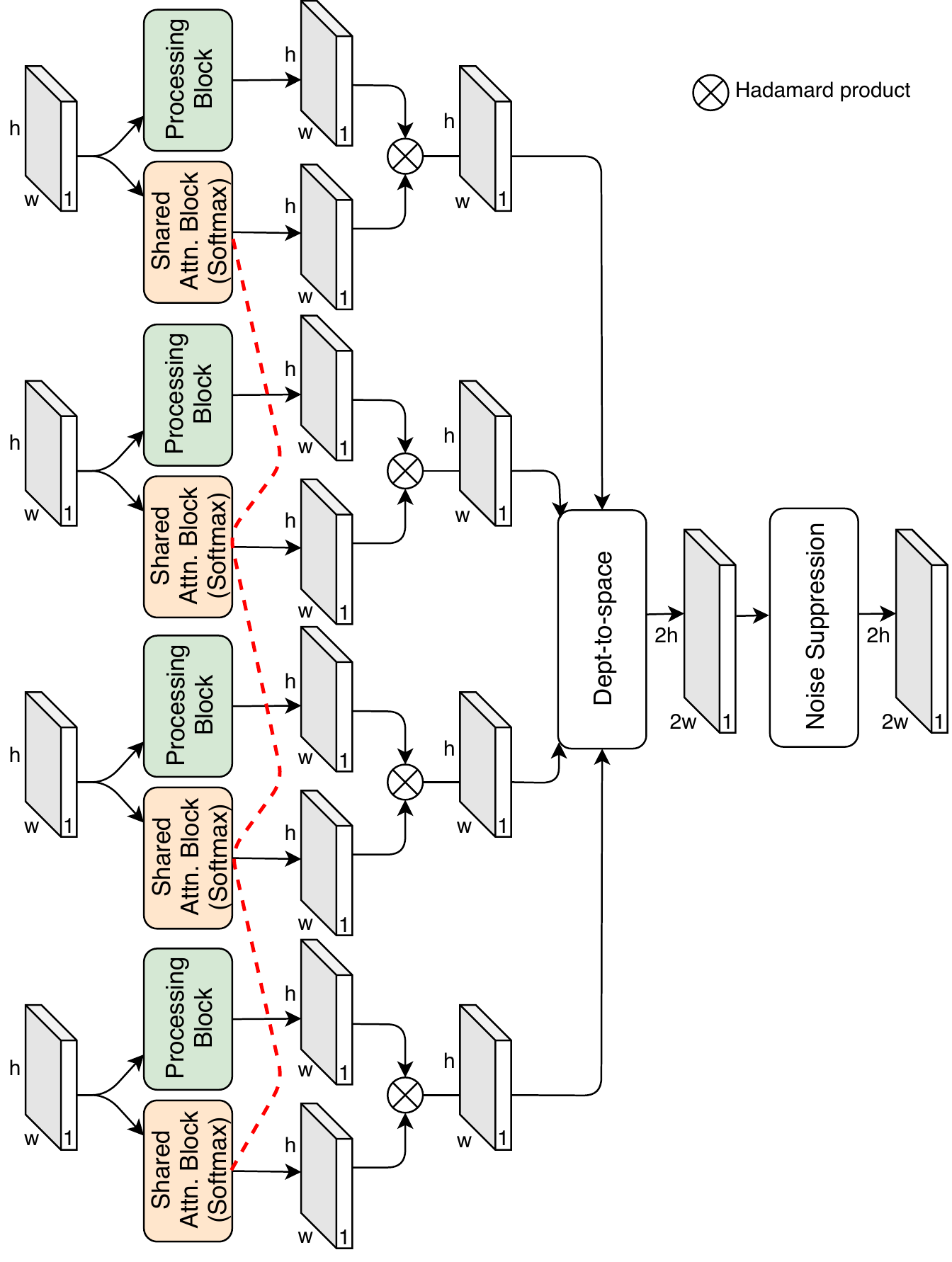}
\caption{Architecture of a 2$\times$ upsampling module that has been used in WDN (as shown in Fig. \ref{fig:model}). The detailed architecture of processing and shared attention block is visualised in Fig. \ref{fig:processing_attention}. The red dotted line represents parameter sharing between blocks.}
\label{fig:network}
\end{figure}

\begin{figure}[t]
\centering
 \includegraphics[width=0.95\linewidth]{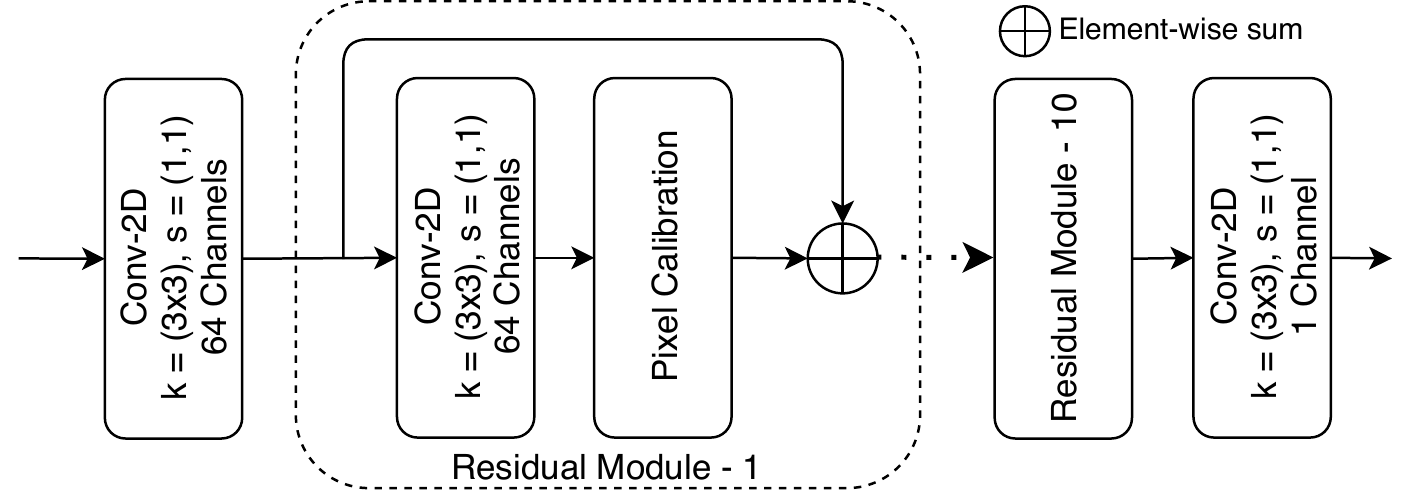}
\caption{Architecture of the processing block and shared attention block as used in the 2$\times$ upsampling module (Fig. \ref{fig:network}) and in the output module (Fig. \ref{fig:output}). The operations of Pixel calibration layer are shown in Eq. \ref{eqn:mhn}}
\label{fig:processing_attention}
\end{figure}

The architecture of a single 2$\times$ upsampling module has been visualised in Fig. \ref{fig:network}. It consists of four processing and four shared attention blocks that process the four inputs (in parallel) that the module receives. Both the processing and shared attention blocks have a similar architecture, as shown in Fig. \ref{fig:processing_attention}. The purpose of a processing block is to improve the quality of its input features, and the purpose of the shared attention block is to decide the relative importance of the features that are generated from different processing blocks. As the shared attention block has to consider all the four inputs to decide their relative importance, the four blocks share the same set of trainable parameters. A Softmax activation is also applied to the outputs of all the four shared attention blocks to normalise the relative importance of pixels at the corresponding locations to one. The processing blocks consider only their respective input, and hence each processing block has its exclusive set of trainable parameters without any sharing. Nevertheless, the output of the processing block gets multiplied with the output of the corresponding attention block and the processed-weighed output generated for the four inputs of the 2$\times$ upsampling module are merged using the depth-to-space operator (shown in Fig. \ref{fig:s2d2}) to generate an upsampled channel of double the input size. The generated channel is convolved with a Gaussian kernel (size = 13$\times$13, $\sigma$ = 0.7) to suppress any undesired noise that might appear. Though this operation induces a little blur in the output, the rest of the network is capable of easily recovering from the minor losses to the high-frequency details caused due to this operation.

\begin{figure}[t]
\centering
 \includegraphics[width=0.85\linewidth]{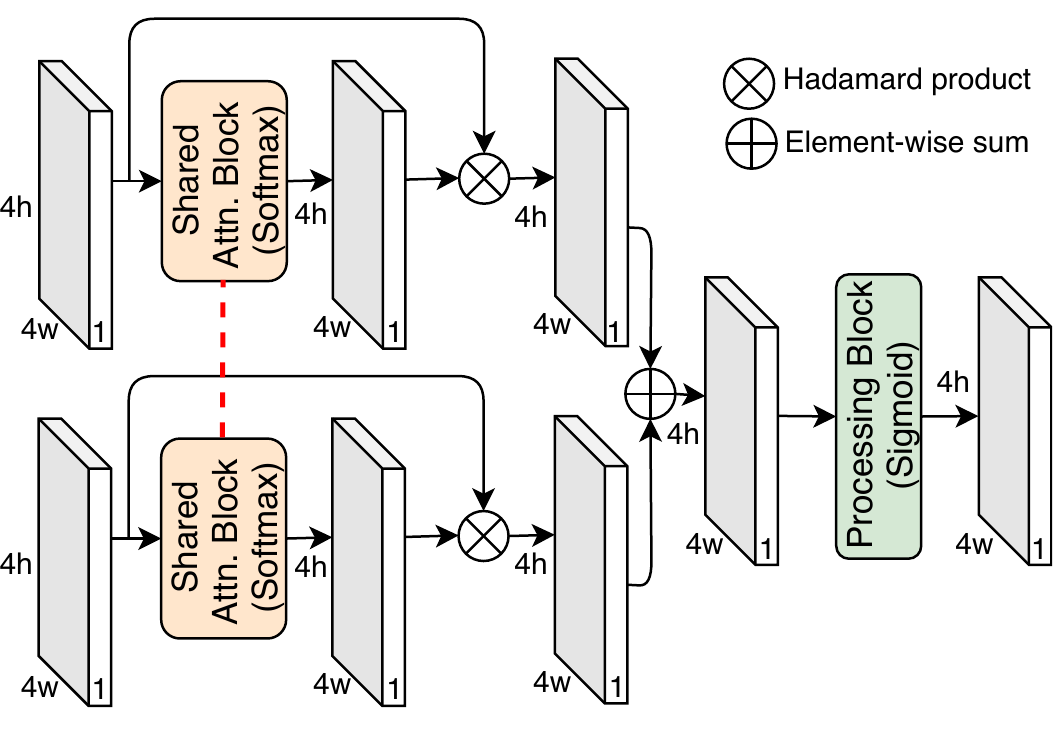}
\caption{Architecture of the output module that has been used in WDN and shown in Fig. \ref{fig:model}. The detailed architecture of the processing and shared attention block is visualised in Fig. \ref{fig:processing_attention}. The red dotted line represents parameter sharing between blocks.}
\label{fig:output}
\end{figure}

The third stage (Ref. Fig. \ref{fig:model}) of WDN consists of an output module, the detailed architecture of which is shown in Fig. \ref{fig:output}. The purpose of the output module is to fuse the upsampled high-frequency and low-frequency channels that are generated by Stage-2 to generate the final upsampled network prediction. This stage accepts as input the two upsampled channels (output of Stage-2) that carry the high-frequency and low-frequency details of the image. First, the stage computes two attention maps, one each for the input high-frequency and low-frequency channels using a `shared attention block' (shown in Fig. \ref{fig:processing_attention}). Next, the attention maps are multiplied with their corresponding high-frequency and low-frequency channels. The resultant products are summed and finally the resultant sum is processed by a `processing block' (shown in Fig. \ref{fig:processing_attention}) to generate the single Luminance channel of the desired output, i.e. the model prediction. 

The motivation for the use of multiple, independent and parallelly connected processing/shared-attention blocks in different stages of WDN is to make each of them expert in solving a particular sub-problem. Moreover, considering that a single processing/shared-attention block takes unit time, the wider design of WDN can speed-up the computation up to 64 times in Stage-1. Stage-1 has eight modules with 64 processing and shared-attention blocks that can process the data in parallel. If these blocks are executed simultaneously on a multi-GPU system, then the said speed-up may be observed as compared to the same 64 blocks connected in sequence. System-level overheads such as data loading and transfer to-and-from GPU will reduce the maximum achievable speed-up.

\textit{Pixel calibration layer:}
As can be seen in the architectural visualisations of our network, WDN often makes use of a layer named pixel calibration. It is a self-learned complex-layer (a layer with many layers) that introduces non-linearity in the network. It replaces the activation function (for instance ReLU) that are typically added after a convolutional layer. Pixel calibration layer in place of this activation function. This layer learns the relative importance of pixels and scales them accordingly. It is a self-learned layer as the learning is performed, from the pixels themselves.

For calibration, a relevance value (between 0-1) is computed for each pixel within a feature map with a Conv2D-Sigmoid operation. Additionally, an irrelevance value is also computed as the difference between relevance value and one. Lastly, positive pixel values are weighed with their respective relevance value, and all the pixels are weighed with their respective irrelevance value and summed up to produce the calibrated output. These operations are shown in Eq. \ref{eqn:mhn}.

\begin{equation}
\label{eqn:mhn}
\begin{split}
Calibrate(y) &= (relu(y)\times V) + (y\times (1.0-V))\\
\quad where \quad V      &=  sigmoid(conv2d(y))
\end{split}
\end{equation}
where the stride of Conv2D is one, and its kernel size is three. The number of output channels generated from Conv2D equals the number of input channels in $y$. $V$ is considered as the relevance value.

This layer's design takes its inspiration from Srivastava \etal \cite{NIPS2015_5850} but is also different. The work of Srivastava \etal has the concept of `transform and carry' gates to train deep networks. The transform gate selects the Convoluted-activated input that is allowed to pass through, and the carry gate selects the actual input that is allowed to pass through the layer. For super-resolution, we adapt the transform and carry concept proposed by Srivastava \etal\ to represent relevance and irrelevance values, respectively. In this adaptation, the concept of `irrelevance/carry' remains the same in WDN as in the related work. However the `relevance/transform' gets changed, in place of transforming the convoluted input feature map $y$ after activation, we transform $y$ after activation directly without convoluting it, to control the relevance of pixels in the feature map. In Eq. \ref{eqn:mhn}, the expression on the left of $+$ represents transform/relevance computation and the expression on the right represents carry/irrelevance computation. Readers are requested to refer Srivastava \etal, to gain a better understanding of transform and carry operations and the stated difference. The performance improvement that is obtained by the stated modification has been shown later in Table \ref{tab:pc_study}.

\subsection{Visualising the intermediate channels/images}
\label{subsec:vam}
To better understand and analyse the architecture of WDN, we visualise all the intermediate channels/images that are generated in Stage-2 and Stage-3 for a given input. As the behaviour of Stage-1 is similar to that of Stage-2, we restrict ourselves to the visualisation of the intermediate channels generated in Stage-2 only.

\begin{figure}[t]
\centering
 \includegraphics[width=00.75\linewidth]{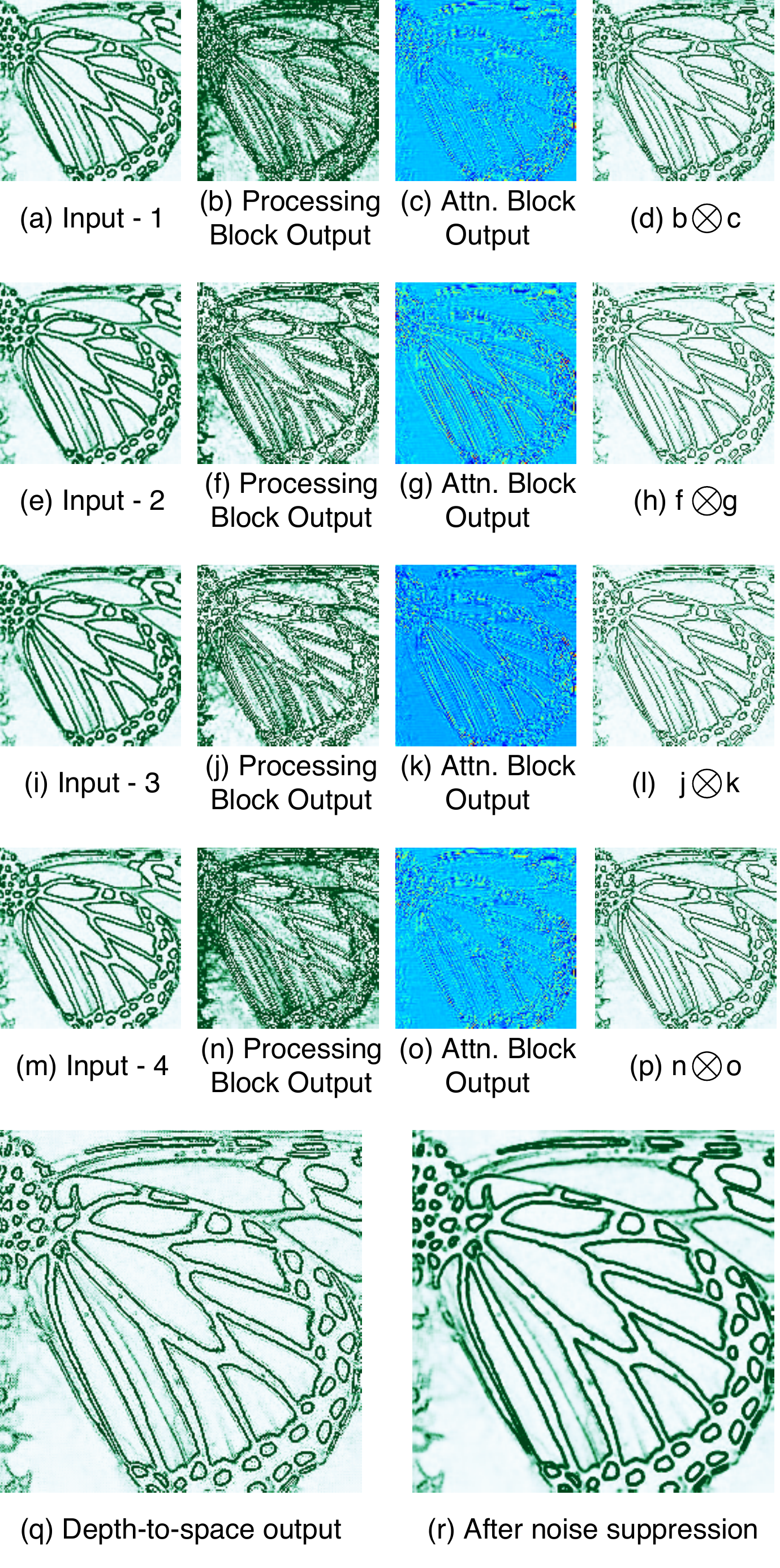}
\caption{Visualisation of intermediate channels generated in Stage-2 for high-frequency input.}
\label{fig:stage-2_hf_plots}
\end{figure}

\begin{figure}[t]
\centering
 \includegraphics[width=0.75\linewidth]{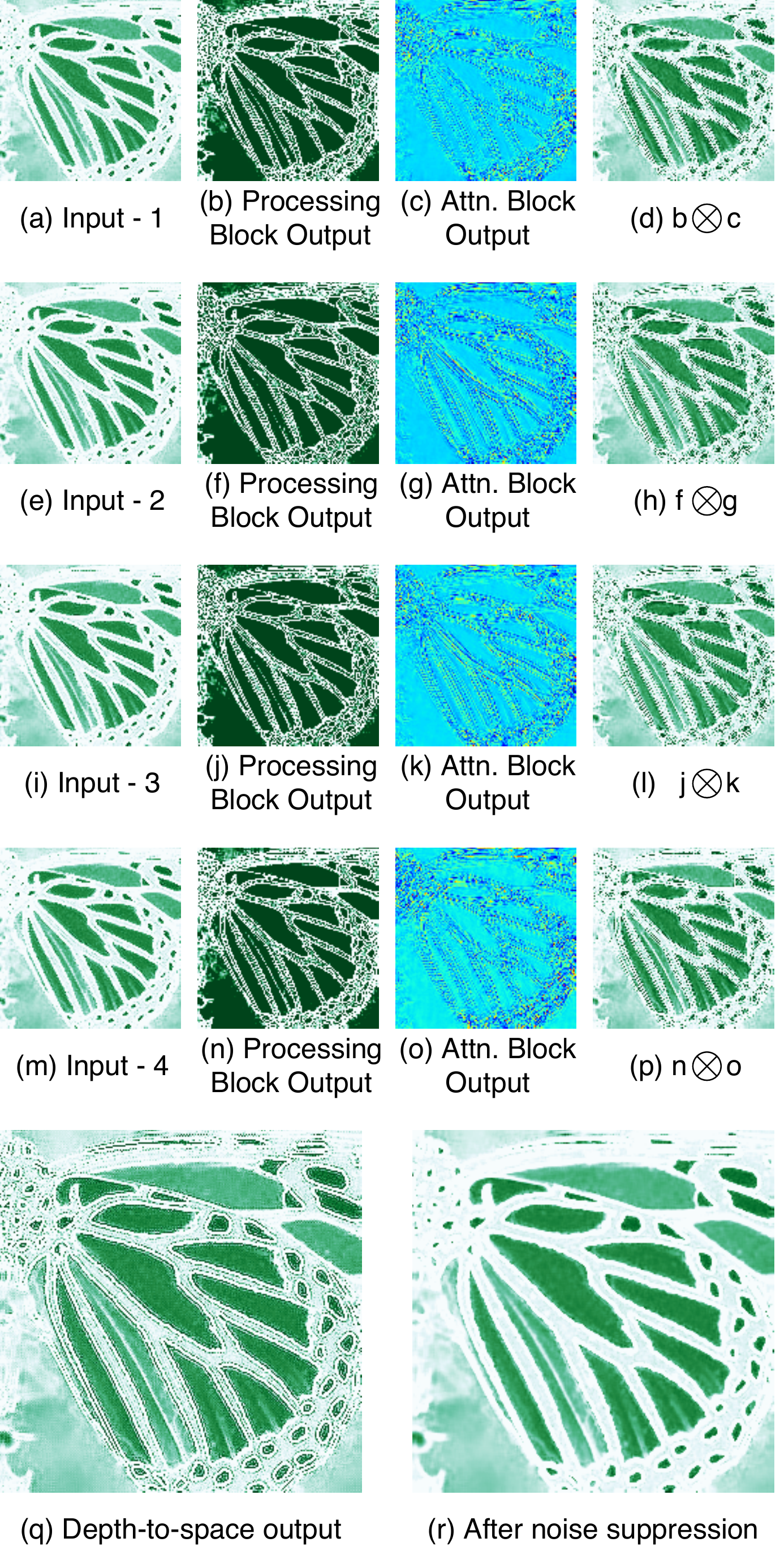}
\caption{Visualisation of intermediate channels generated in Stage-2 for low-frequency input.}
\label{fig:stage-2_lf_plots}
\end{figure}

\begin{figure}[t]
\centering
 \includegraphics[width=0.95\linewidth]{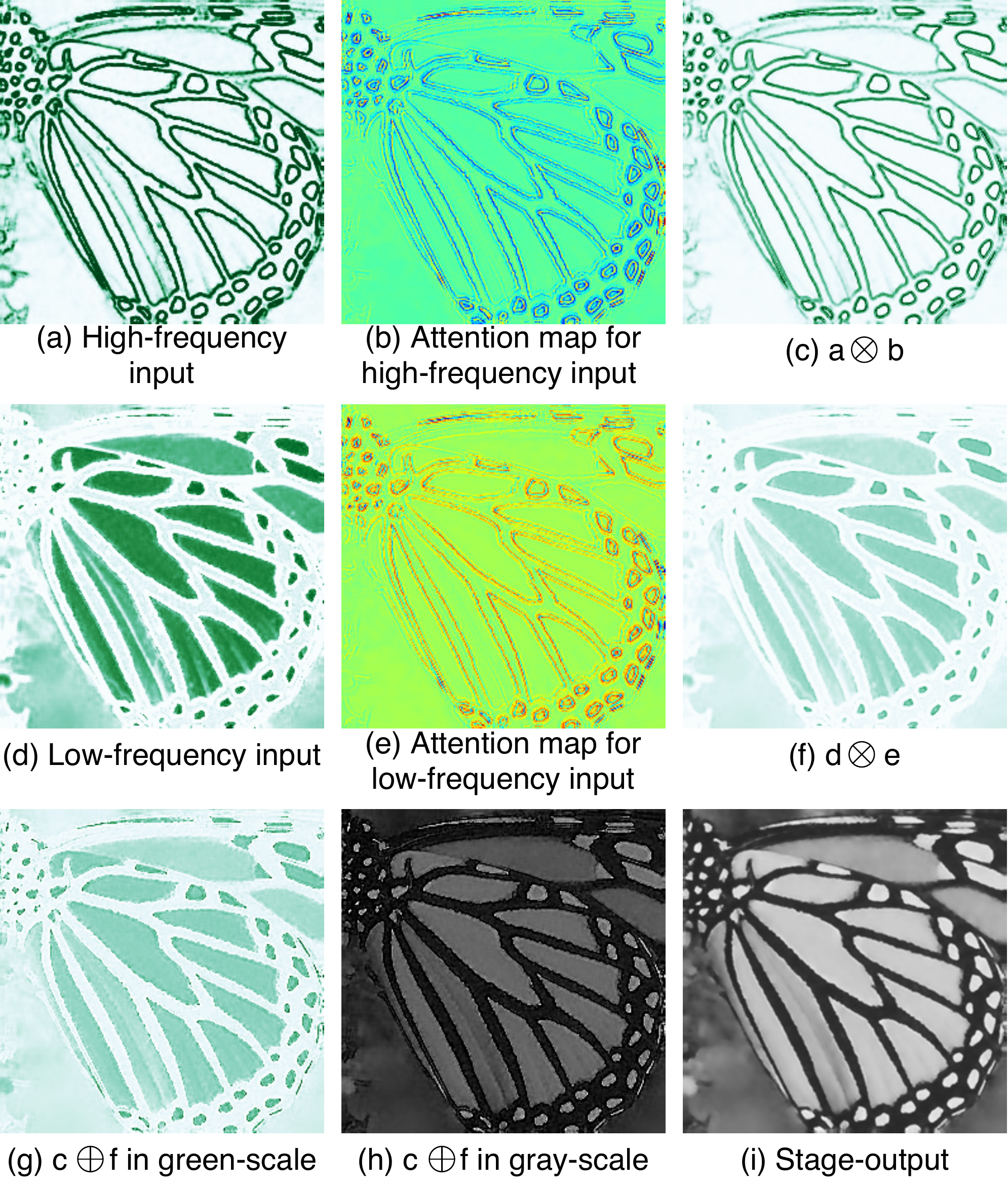}
\caption{Visualisations of intermediate channels generated in Stage-3.}
\label{fig:output_module_plots}
\end{figure}

Figs. \ref{fig:stage-2_hf_plots} and \ref{fig:stage-2_lf_plots} visualise the intermediate channels that are generated in Stage-2 for high-frequency and low-frequency inputs respectively, while, Fig. \ref{fig:output_module_plots} visualises the intermediate channels of Stage-3. It can be seen in the figures that each block that has been used in the architecture of WDN performs its task as per its design objective. The processing blocks and shared attention blocks jointly improve the quality of input features, depth-to-space increases the resolution by two, and the Gaussian noise suppressor suppresses the noise to generate an upsampled channel of better quality.

\subsection{Training losses and ground-truth}
The procedure to generate the ground-truth for all the sub-problems that WDN attempts to solve were described in Section \ref{subsec:subprb}. We now describe the losses that are to be minimised to train WDN. The first and second stages of WDN that comprise of 2$\times$ upsampling modules are trained by minimising the Mean-Square-Error (MSE as shown in Eq. \ref{eqn:mses1}) between the stage predictions and the corresponding ground-truths.

\begin{equation}
Loss_{upsampling} = \frac{\sum\limits_{i=1}^{n}{(y_{i}-y'_{i})^2}}{n}
\label{eqn:mses1}
\end{equation}

\noindent where $n$ is a scalar having the value equivalent to the number of pixels in the ground-truth, $y_i$ is the module ground-truth, and $y'_i$ is the module prediction. Similarly, the third stage that has an output module is trained by minimising the loss (shown in Eq. \ref{eqn:mses2}) between the stage prediction and the corresponding ground-truth:

\begin{equation}
Loss_{output} = MSE(y,y') + (1-SSIM(y,y'))
\label{eqn:mses2}
\end{equation}

\noindent where $y$ is the model ground-truth and $y'$ is the model prediction. Mean-square-error minimisation maximises the PSNR \cite{IRANI1993324} metric while 1-SSIM (structural dissimilarity) minimisation maximises the SSIM \cite{1284395} metric.

\begin{table*}
\footnotesize
\caption{Quantitative comparison (4$\times$) with state-of-the-art methods. Evaluation procedure used to compute the values is as used/described by Qiu \etal \cite{Qiu_2019_ICCV}. All values have been captured from the non-self-ensemble variant. \textbf{Note: We plan to release our codes upon acceptance of our work.}}
\label{tab:isrquantitative}
\centering
\begin{tabular}{ c c | c c | c c | c c | c c | c c}
& & \multicolumn{2}{c|}{Set5} & \multicolumn{2}{c|}{Set14} & \multicolumn{2}{c}{B100} & \multicolumn{2}{c}{Urban100} & \multicolumn{2}{c}{Manga109}\\
\hline
Method & Model & PSNR & SSIM & PSNR & SSIM & PSNR & SSIM & PSNR & SSIM & PSNR & SSIM\\
\hline
\hline
Bicubic & Bicubic & 28.42 & 0.8104 & 26.00 & 0.7027 & 25.96 & 0.6675 & 23.14 & 0.6577 & 24.89 & 0.7866\\
\hline
Lim \etal (2017) \cite{8014885} & EDSR & 32.46 & 0.8968 & 28.80 & 0.7876 & 27.71 & 0.7420 & 26.64 & 0.8033 & 31.02 & 0.9148 \\
\hline
Haris \etal (2018) \cite{8578277} & D-DBPN & 32.47 & 0.8980 & 28.82 & 0.7860 & 27.72 & 0.7400 & 26.38 & 0.7946 & 30.91 & 0.9137\\
\hline
Li \etal (2019)\cite{Li_2019_CVPR} & SRFBN & 32.47 & 0.8983 & 28.81 & 0.7868 & 27.72 & 0.7409 & 26.60 & 0.8015 & 31.15 & 0.9160\\
\hline
Zhang \etal (2018) \cite{8578360} & RDN & 32.47 & 0.8990 & 28.81 & 0.7871 & 27.72 & 0.7419 & 26.61 & 0.8028 & 31.00 & 0.9151\\
\hline
Lim \etal (2017) \cite{8014885} & MDSR & 32.50 & 0.8973 & 28.72 & 0.7857 & 27.72 & 0.7418 & 26.67 & 0.8041 & - & - \\
\hline
He \etal (2019) \cite{He_2019_CVPR} & OISR-RK3 & 32.53 & 0.8992 & 28.86 & 0.7878 & 27.75 & 0.7428 & 26.79 & 0.8068 & - & - \\
\hline
Qin \etal (2020) \cite{Qin2020} & MRAN & 32.61 & 0.8998 & 28.82 & 0.7875 & 27.73 & 0.7420 & 26.70 & 0.8051 & 30.92 & 0.9147\\
\hline
Zhang \etal (2018) \cite{10.1007/978-3-030-01234-2_18} & RCAN & 32.63 & 0.9002 & 28.87 & 0.7889 & 27.77 & 0.7436 & 26.82 & 0.8087 & 31.22 & 0.9173\\
\hline
Dai \etal (2019) \cite{Dai_2019_CVPR} & SAN & 32.64 & 0.9003 & 28.92 & 0.7888 & 27.78 & 0.7436 & 26.79 & 0.8068 & 31.18 & 0.9169\\
\hline
Qiu \etal (2019) \cite{Qiu_2019_ICCV} & EBRN & 32.79 & 0.9032 & 29.01 & 0.7903 & 27.85 & 0.7464 & 27.03 & 0.8114 & 31.53 & 0.9198\\
\hline
\textbf{Ours} & \textbf{WDN} & \textbf{33.10} & \textbf{0.9092} & \textbf{29.21} & \textbf{0.7929} & \textbf{27.98} & \textbf{0.7519} & \textbf{27.51} & \textbf{0.8197} & \textbf{32.17} & \textbf{0.9247}\\
\hline
\end{tabular}
\end{table*}

\subsection{Training details}
To sum up the training process, 32 inputs of Stage-1 are generated by applying space-to-depth (block size: four) on bicubically upsampled and frequency-separated given low-resolution input. Eight inputs of Stage-2 are generated in the form of Stage-1 output. The stage-loss for both Stage 1 and 2 is shown in Eq. \ref{eqn:mses1}. Two inputs of Stage-3 are generated in the form of Stage-2 output, and the stage-loss is shown in Eq. \ref{eqn:mses2}. Each stage requires separate training with inputs that are generated by the trained previous stage. The procedure to generate ground-truths for training all the three stages has been discussed in Section \ref{subsec:subprb}.

All trainable weights are initialised with the default \textit{Glorot initialisation} \cite{pmlr-v9-glorot10a}. Adam \cite{DBLP:journals/corr/KingmaB14}, with a fixed learning rate = $10^{-4}$, $\beta_1$ = 0.9, $\beta_2$ = 0.999, and $\epsilon$ = 1e-08, has been used to optimise the training. All the stages have been trained one-by-one after freezing the parameters of the previous stage. Wherever required, reflective padding has been applied in the appropriate dimensions of the input of all the Convolutional layers. In order to make sure that the network does not over-fit on the training data, standard countermeasures such as data augmentation (random cropping, rotation and horizontal flipping), regularisation and early stopping have been enforced. Training of a stage is considered as complete when no significant improvement reflects in the performance metric on validation data for five consecutive epochs. The models have been trained and tested on Google's Tensor Processing Unit (TPU) and V100 GPUs.

\section{Experiments and analysis}
\label{sec:experiment}
Extensive experiments have been performed on multiple datasets to evaluate the efficacy of the ideas and architecture proposed in this work. Typically, Peak Signal-to-Noise Ratio (PSNR \cite{IRANI1993324}), and Structural Similarity Index (SSIM \cite{1284395}) are the metrics that are used to quantify the performance of a super-resolution technique and compare it with others. Hence, the same have been used in this work also. We now describe the datasets that have been used in the experiments to train, validate and evaluate WDN. 

Similar to the methods \cite{8014885,8578442,8578360,10.1007/978-3-030-01234-2_18,Qiu_2019_ICCV}, WDN is also trained on the DIV2K dataset by Timofte \etal \cite{8014883}. DIV2K dataset contains 1000 images at 2K resolution, among which 800 are for training, and 100 are for validation. The evaluation procedure and other experimental settings (unless stated explicitly) have been adopted from the current state-of-the-art method by Qiu \etal \cite{Qiu_2019_ICCV}. The comparison of WDN with the existing benchmarks has been made on five publicly available datasets, namely Set5 \cite{set5,6751349}, Set14 \cite{set14}, B100 \cite{937655,10.1007/978-3-319-16817-3_8}, Urban100 \cite{7299156}, and Manga109 \cite{Matsui2017}. 

\begin{figure*}
\centering
\begin{subfigure}{0.19\textwidth}
 \centering
 \includegraphics[width=0.99\linewidth]{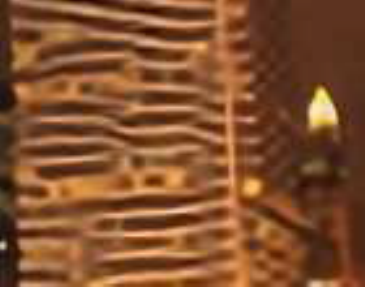}
 \caption{\cite{7780551}}
 \label{fig:2102}
\end{subfigure}%
\begin{subfigure}{0.19\textwidth}
 \centering
 \includegraphics[width=0.99\linewidth]{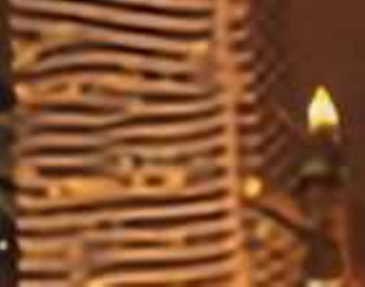}
 \caption{\cite{7780550}}
 \label{fig:2103}
\end{subfigure}%
\begin{subfigure}{0.19\textwidth}
 \centering
 \includegraphics[width=0.99\linewidth]{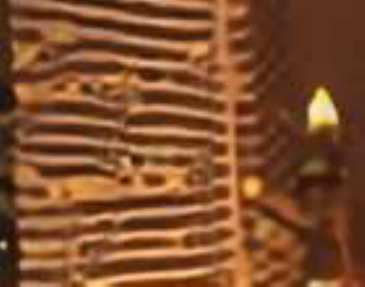}
 \caption{\cite{8100101}}
 \label{fig:2104}
\end{subfigure}%
\begin{subfigure}{0.19\textwidth}
 \centering
 \includegraphics[width=0.99\linewidth]{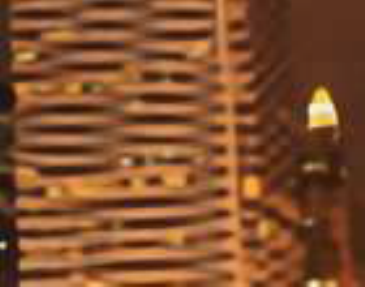}
 \caption{\cite{8014885}}
 \label{fig:2105}
\end{subfigure}%
\begin{subfigure}{0.19\textwidth}
 \centering
 \includegraphics[width=0.99\linewidth]{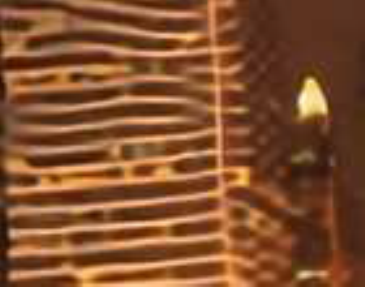}
 \caption{\cite{8578180}}
 \label{fig:2106}
\end{subfigure}

\begin{subfigure}{0.19\textwidth}
 \centering
 \includegraphics[width=0.99\linewidth]{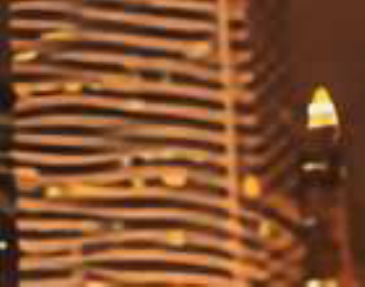}
 \caption{\cite{10.1007/978-3-030-01237-3_32}}
 \label{fig:2107}
\end{subfigure}%
\begin{subfigure}{0.19\textwidth}
 \centering
 \includegraphics[width=0.99\linewidth]{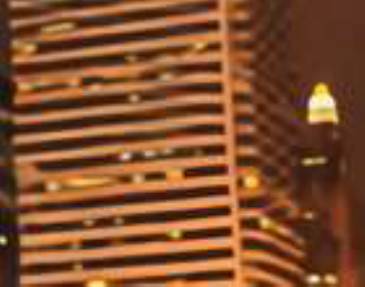}
 \caption{\cite{Qiu_2019_ICCV}}
 \label{fig:2108}
\end{subfigure}%
\begin{subfigure}{0.19\textwidth}
 \centering
 \includegraphics[width=0.99\linewidth]{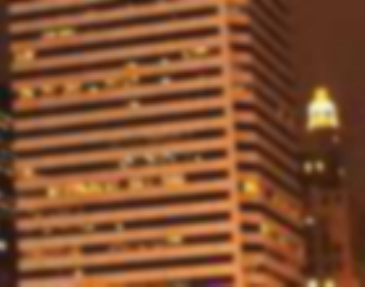}
 \caption{\textbf{WDN}}
 \label{fig:2109}
\end{subfigure}%
\begin{subfigure}{0.19\textwidth}
 \centering
 \includegraphics[width=0.99\linewidth]{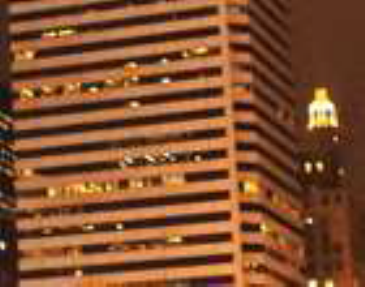}
 \caption{Ground truth}
 \label{fig:2110}
\end{subfigure}%
\begin{subfigure}{0.19\textwidth}
 \centering
 \includegraphics[width=0.99\linewidth]{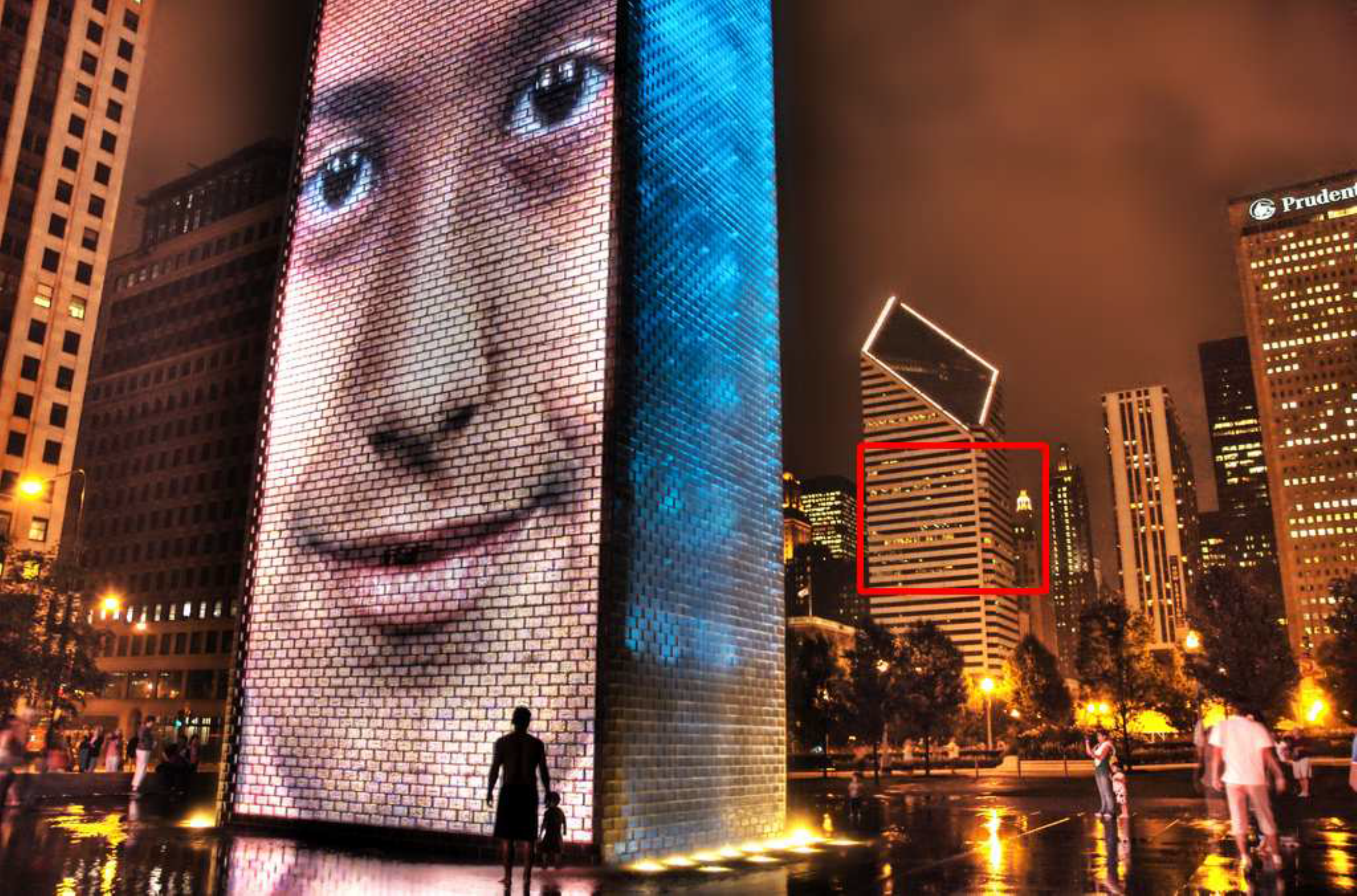}
 \caption{GT full}
 \label{fig:2111}
\end{subfigure}

\begin{subfigure}{0.19\textwidth}
 \centering
 \includegraphics[width=0.99\linewidth]{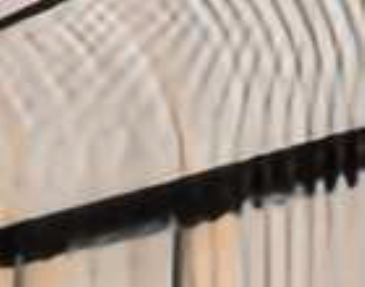}
 \caption{\cite{7780551}}
 \label{fig:2002}
\end{subfigure}%
\begin{subfigure}{0.19\textwidth}
 \centering
 \includegraphics[width=0.99\linewidth]{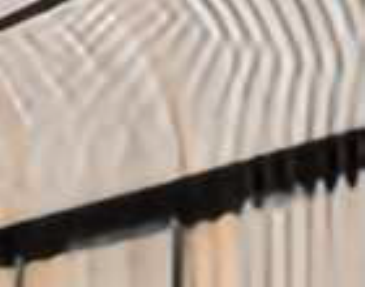}
 \caption{\cite{7780550}}
 \label{fig:2003}
\end{subfigure}%
\begin{subfigure}{0.19\textwidth}
 \centering
 \includegraphics[width=0.99\linewidth]{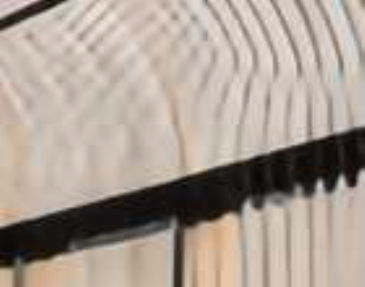}
 \caption{\cite{8100101}}
 \label{fig:2004}
\end{subfigure}%
\begin{subfigure}{0.19\textwidth}
 \centering
 \includegraphics[width=0.99\linewidth]{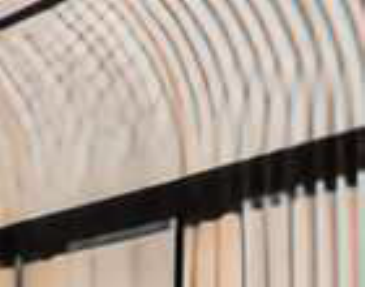}
 \caption{\cite{8014885}}
 \label{fig:2005}
\end{subfigure}%
\begin{subfigure}{0.19\textwidth}
 \centering
 \includegraphics[width=0.99\linewidth]{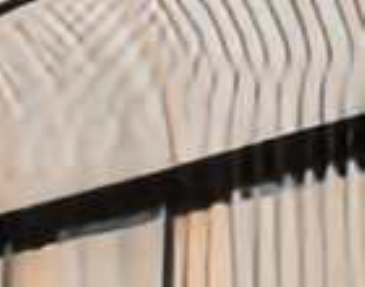}
 \caption{\cite{8578180}}
 \label{fig:2006}
\end{subfigure}

\begin{subfigure}{0.19\textwidth}
 \centering
 \includegraphics[width=0.99\linewidth]{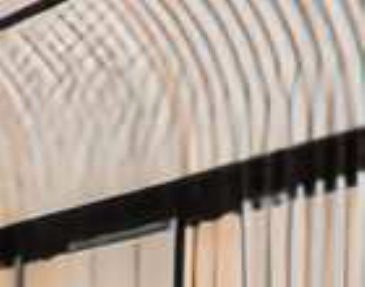}
 \caption{\cite{10.1007/978-3-030-01237-3_32}}
 \label{fig:2007}
\end{subfigure}%
\begin{subfigure}{0.19\textwidth}
 \centering
 \includegraphics[width=0.99\linewidth]{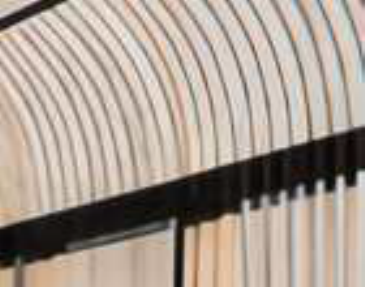}
 \caption{\cite{Qiu_2019_ICCV}}
 \label{fig:2008}
\end{subfigure}%
\begin{subfigure}{0.19\textwidth}
 \centering
 \includegraphics[width=0.99\linewidth]{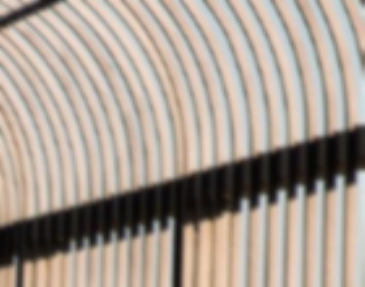}
 \caption{\textbf{WDN}}
 \label{fig:2009}
\end{subfigure}%
\begin{subfigure}{0.19\textwidth}
 \centering
 \includegraphics[width=0.99\linewidth]{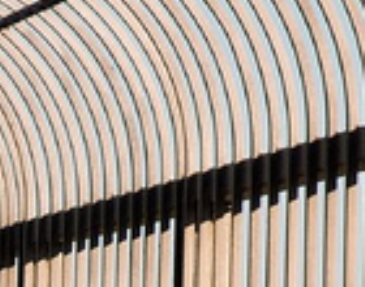}
 \caption{Ground truth}
 \label{fig:2010a}
\end{subfigure}%
\begin{subfigure}{0.19\textwidth}
 \centering
 \includegraphics[width=0.99\linewidth]{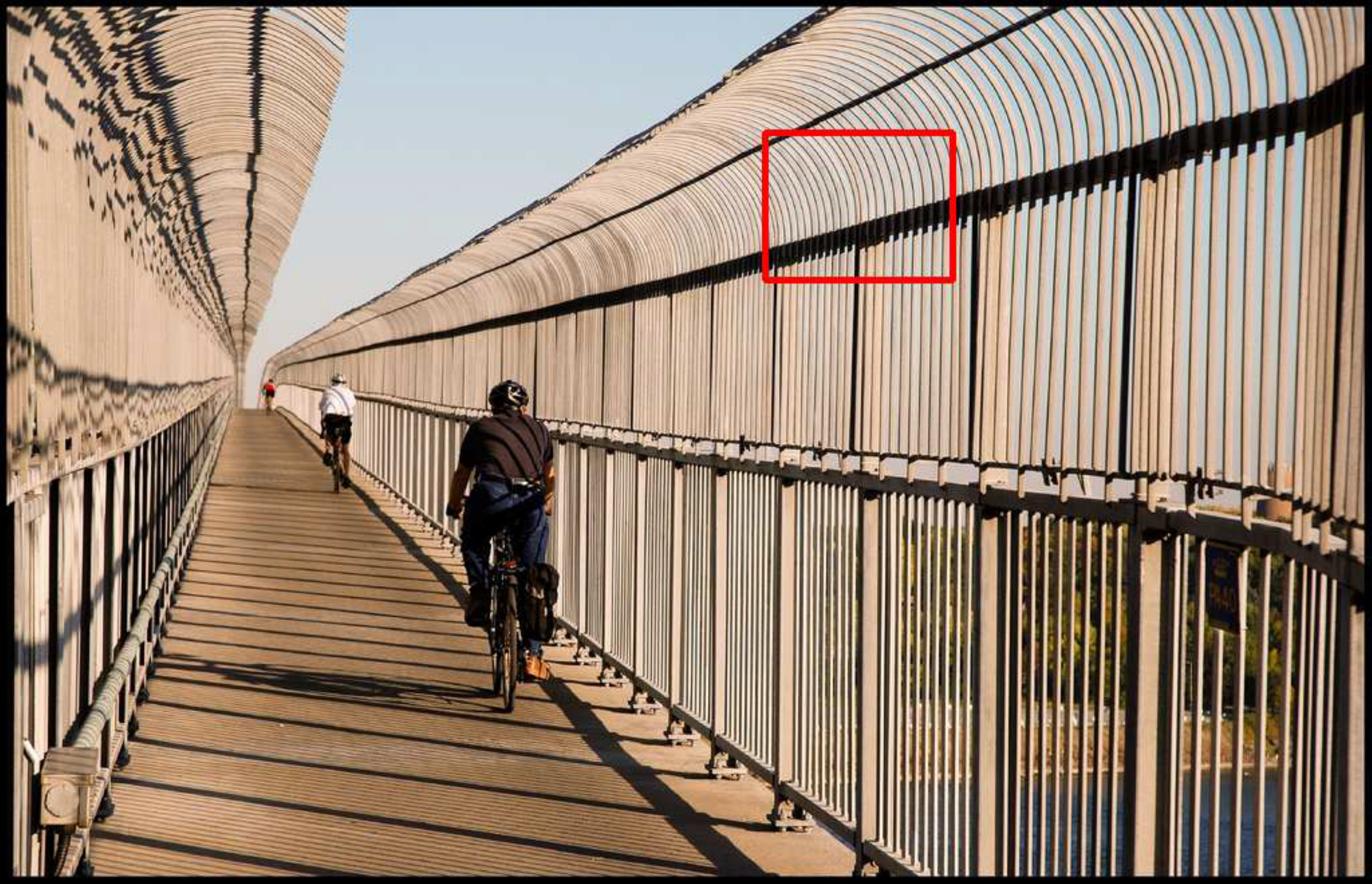}
 \caption{GT full}
 \label{fig:2011a}
\end{subfigure}
\caption{Predicted image visualisation for qualitative comparison on two images from the Urban100 dataset \cite{7299156} for 4$\times$ super-resolution.}
\label{fig:qualisr1}
\end{figure*}

\subsection{Comparison with the state-of-the-art}
\label{subsec:sota}
Table \ref{tab:isrquantitative} presents the quantitative results and Fig. \ref{fig:qualisr1}  visualises the upsampled images on a scale factor of four for comparison with the state-of-the-art methods. It can be seen that the results generated by WDN are numerically better, visually sharper, and visually less noisy than the existing techniques. For this improved performance of WDN, we credit to the: 1) Divide-and-conquer design paradigm, 2) Wide and deep network design, 3) Explicit high-frequency channel prediction, and 4) Pixel calibration layer.

\textit{Performance on scaling factors of two and three:}
This computation requires the following modification in WDN: For a scale of two, we assume that the given input low-resolution image already has the dimension 2h$\times$2w$\times$1, and hence we modify the bi-cubic upsampling (refer Fig. \ref{fig:model}) to upsample 2$\times$ instead of 4$\times$. Similarly, for a scale of three, the bicubic upsampling is modified to upsample 3$\times$. Rest of the network architecture remains the same. The results obtained on the upsampling scales of two and three are shown in Table \ref{tab:other_scale}. It can be seen that WDN performs better than the cited methods. The possible reasons for this improvement are the same as mentioned in the last paragraph.

\subsection{Analysing the parameters and computation time}
\label{subsec:pct}
\begin{figure}
\centering
\includegraphics[width=0.97\linewidth]{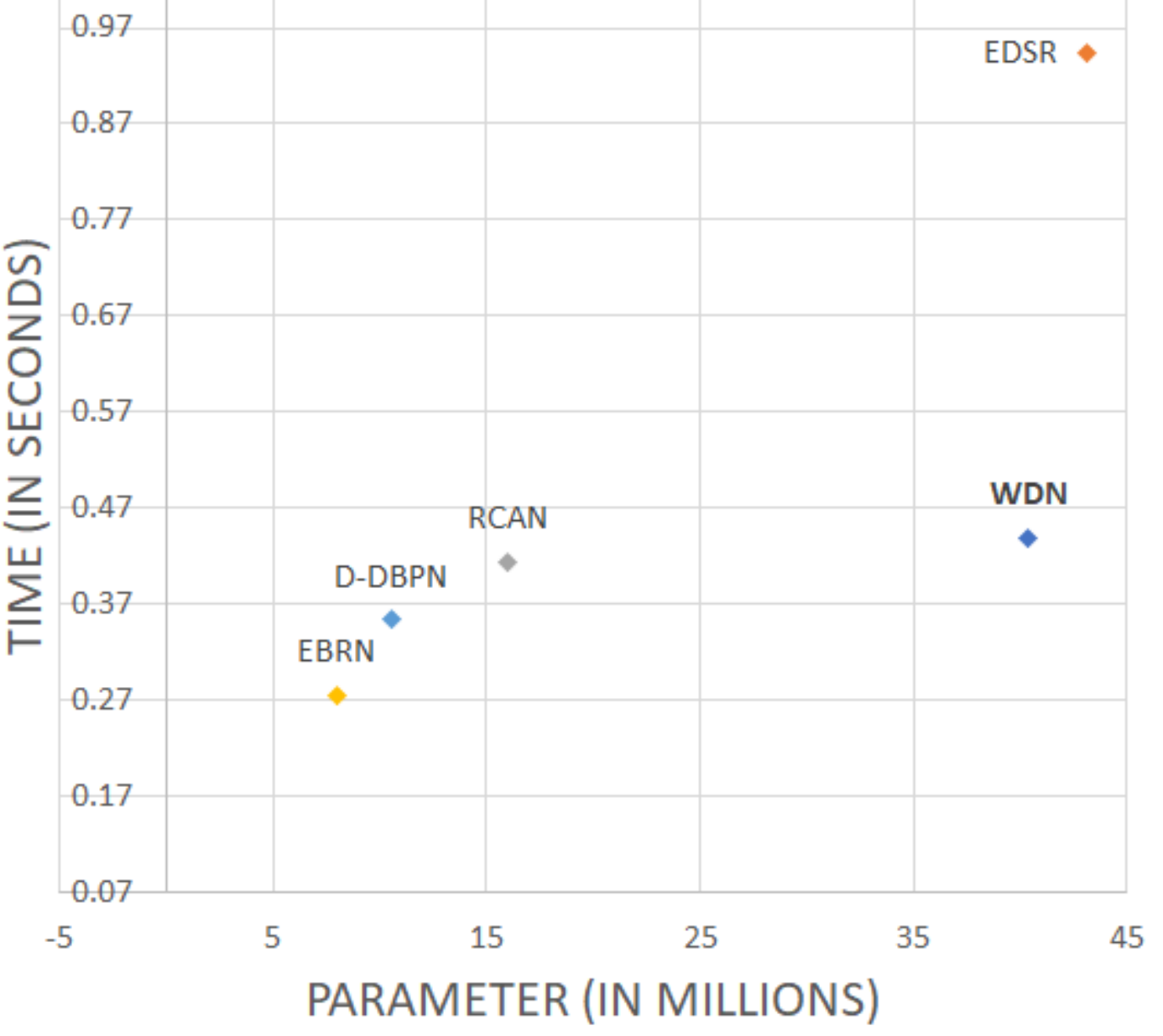}
\caption{Graph showing the computation time vs the number of trainable parameters in state-of-the-art networks and WDN.}
\label{fig:ptime}
\end{figure}

To further analyse the complexity of WDN, we show the total number of parameters, floating-point operations (FLOPS) and processing time of each stage of WDN in Table \ref{tab:pct}. We also compare the parameters and processing time of WDN with a few state-of-the-art networks in Fig. \ref{fig:ptime}.

\begin{table}[t]
\footnotesize
\caption{Number of trainable parameters, floating-point operations (FLOPS), and processing time of each stage in WDN. Parameters and FLOPS are in millions. Time is in seconds.}
\label{tab:pct}
\centering
\begin{tabular}{ c || c c c c c  }
\hline
\textit{Stage $\rightarrow$} & Preproc. & 1 & 2 & 3 & Total\\
\hline
Parameters & - & 31.07 & 7.77 & 1.55 & 40.39\\
\hline
FLOPS & - & 62.02 & 15.51 & 3.10 & 80.63\\
\hline
Time & 0.0103 & 0.1421 & 0.2234 & 0.0722 & 0.4390\\
\hline
\end{tabular}
\end{table}

It can be seen in Table \ref{tab:pct} that 1) Stage-1 has the maximum number of parameters; this is because it has to process 32 inputs, 2) Stage-2 has lesser parameters as it processes only eight inputs, and 3) Stage-3 has the least number of parameters as it processes only two input channels. The floating-point operations show a similar trend as the number of parameters. In terms of the processing time in each stage, it can be seen that Stage-2 is slightly slower than Stage-1 despite having a lesser number of parameters. Unlike Stage-1, Stage-2 has to process 2$\times$ upsampled channels that are generated by Stage-1, and due to this, the processing time of Stage-2 becomes slightly more than that of Stage-1. Stage-3 processes 4$\times$ upsampled images; however, due to a lesser number of FLOPS, this stage has the fastest processing time. The recent hardware developments have resulted in GPUs/TPU with large memory at a much cheaper cost, and so parametric heaviness should not become an obstacle in the application of WDN when accuracy has the priority over memory constraint. We also mention that: 1) The total processing time of individual stages is more than the processing time of the full network due to system-level overheads involved in individually computing the time for each stage. 2) The processing time has been evaluated with maximum parallelism that can be attained using multiple GPUs/TPU.

It can be seen in Fig. \ref{fig:ptime} that WDN is heavier in terms of parameters than some state-of-the-art networks, but its processing time is comparable to other networks. This is primarily due to the wide design of WDN, that can take advantage of the multiple GPUs/TPU for parallel processing. Nevertheless, restoration tasks such as super-resolution are typically not constrained with a real-time response requirement, and so processing time or larger number of parameters should not become a hurdle in the application of WDN when prediction quality has the priority over the processing time. The time measurements were obtained on an n1-standard-4 system with Nvidia Tesla V100 GPU on GCP. An image of size 1980 $\times$ 1080 was used for 4$\times$ upsampling. The parameter values have been obtained from the respective publications, and the time values have been computed. The actual time may vary due to the library-specific optimisations; dynamic shared system load and GPU thermal slowdown in place during the measurement.

\begin{table}[t]
\footnotesize
\caption{Quantitative comparison with state-of-the-art methods on a scaling factors of two and three.}
\label{tab:other_scale}
\begin{subtable}[t]{0.47\textwidth}
\fontsize{8.5}{10.5}\selectfont
\caption{Comparison on a scaling factor of two.}
\label{tab:2xpsnr}
\centering
\begin{tabular}{ c|| c c c c c}
\hline
\textit{\textbf{Dataset}} & Metric & SAN & RCAN & EBRN & \textbf{WDN} \\
\hline
\multirow{2}{*}{Set5} & PSNR & 38.31 & 38.27 & 38.35 & \textbf{38.41}\\
\cline{2-6}
& SSIM & 0.9620 & 0.9614 & 0.9620 & \textbf{0.9623}\\
\hline
\multirow{2}{*}{Set14} & PSNR & 34.07 & 34.12 & 34.24 &  \textbf{34.37}\\
\cline{2-6}
& SSIM & 0.9213 & 0.9216 & 0.9226 &  \textbf{0.9234}\\
\hline
\multirow{2}{*}{B100} & PSNR & 32.42 & 32.41 & 32.47 & \textbf{32.50}\\
\cline{2-6}
& SSIM & 0.9028 & 0.9027 & 0.9033 &  \textbf{0.9039}\\
\hline 
\multirow{2}{*}{Urban100} & PSNR & 33.10 & 33.34 & 33.52 & \textbf{33.71}\\
\cline{2-6}
& SSIM & 0.9370 & 0.9384 & 0.9402 & \textbf{0.9421}\\
\hline 
\multirow{2}{*}{Manga109} & PSNR & 39.32 & 39.44 & 39.62 & \textbf{39.81}\\
\cline{2-6}
& SSIM & 0.9792 & 0.9786 & 0.9802 &  \textbf{0.9811}\\
\hline 
\end{tabular}
\end{subtable}

\begin{subtable}[t]{0.47\textwidth}
\fontsize{8.5}{10.5}\selectfont
\vspace{4mm}
\caption{Comparison on a scaling factor of three.}
\label{tab:3xpsnr}
\centering
\begin{tabular}{ c|| c c c c c}
\hline
\textit{\textbf{Dataset}} & Metric & OISR-RK3 & SAN & RCAN & \textbf{WDN} \\
\hline
\multirow{2}{*}{Set5} & PSNR & 34.72 & 34.75 & 34.74 &  \textbf{34.95}\\
\cline{2-6}
& SSIM & 0.9297 & 0.9300 & 0.9299 & \textbf{0.9331}\\
\hline
\multirow{2}{*}{Set14} & PSNR & 30.57 & 30.59 & 30.65 &  \textbf{30.87}\\
\cline{2-6}
& SSIM & 0.8470 & 0.8476 & 0.8482 & \textbf{0.8502}\\
\hline
\multirow{2}{*}{B100} & PSNR & 29.29 & 29.33 & 29.32 & \textbf{29.41}\\
\cline{2-6}
& SSIM & 0.8103 & 0.8112 & 0.8111 & \textbf{0.8151}\\
\hline 
\multirow{2}{*}{Urban100} & PSNR & 28.95 & 28.93 & 29.09 & \textbf{29.49}\\
\cline{2-6}
& SSIM & 0.8680 & 0.8671 & 0.8702 & \textbf{0.8761}\\
\hline 
\multirow{2}{*}{Manga109} & PSNR & - & 34.30 & 34.44 & \textbf{34.96}\\
\cline{2-6}
& SSIM & - & 0.9494 & 0.9499 & \textbf{0.9531}\\
\hline 
\end{tabular}
\end{subtable}
\end{table}

\subsection{Ablation Studies}
\label{subsec:ablation}
\subsubsection{Evaluating different training procedures}
\label{subsec:training}
WDN has been trained stage-by-stage with multiple losses. However, WDN can also be trained by: 1) Minimising all the losses together with no inter-stage gradient flow. 2) Allowing the inter-stage gradient flow with existing losses, and  3) Allowing the inter-stage gradient flow with losses after the last stage only, i.e. end-to-end training.

The first procedure is equivalent to the existing training procedure, the only difference being that the entire model stays in memory during training. Also, in this procedure, the training of the latter stages start to show convergence when the parameters of the previous stage stabilise. Without the stabilisation of the previous stage parameters, the input to latter stages frequently changes, thus delaying its convergence. Table \ref{tab:training} shows the results obtained on different test datasets after following the latter two procedures.

\begin{table}[t]
\centering
\footnotesize
\caption{Results obtained on test datasets upon following different training procedures (Proc.) as described in Section \ref{subsec:training}.}
\label{tab:training}
\begin{tabular}{ c|| c c c c }
 \hline
  \textit{\textbf{Dataset}} & Metric & Proc. 2 & Proc. 3 & Existing/Proc. 1 \\
 \hline
 \multirow{2}{*}{Set5} & PSNR & 32.54 & 30.51 & \textbf{33.10}\\
 \cline{2-5}
 & SSIM & 0.8995 & 0.8635 & \textbf{0.9092}\\
 \hline
 \multirow{2}{*}{Set14} & PSNR & 28.84 & 27.48 & \textbf{29.21}\\
 \cline{2-5}
 & SSIM & 0.7881 & 0.7509 & \textbf{0.7929}\\
 \hline
 \multirow{2}{*}{B100} & PSNR & 27.72 & 26.92 & \textbf{27.98}\\
 \cline{2-5}
 & SSIM & 0.7432 & 0.7097 & \textbf{0.7519}\\
 \hline 
 \multirow{2}{*}{Urban100} & PSNR & 26.82 & 24.55 & \textbf{27.51}\\
 \cline{2-5}
 & SSIM & 0.8071 & 0.7227 & \textbf{0.8197}\\
 \hline 
 \multirow{2}{*}{Manga109} & PSNR & 31.16 & 27.53 & \textbf{32.17}\\
 \cline{2-5}
 & SSIM & 0.9165 & 0.8549 & \textbf{0.9247}\\
 \hline 
 \end{tabular}
\end{table}

It can be inferred from the Table that the second procedure shows some drop in the performance. This might be due to a trainable variable's value getting disturbed/fluctuated due to the influence of multiple losses on it simultaneously. Different losses have different objectives, and they might influence a variable for their respective minimisation. Though not always, but in the current case, this has led the model to underperform. Tuning the $\lambda$s for different losses might be useful here. However, allowing the inter-stage gradient flow would require the whole model to fit in a single GPU and would prevent parallel training.

The third procedure further deteriorated the results, possibly as the wide and deep architecture of WDN has been designed to work by dividing the problem into sub-problems effectively. An altogether different architectural design might be fruitful in this case where end-to-end training is a hard constraint.

\begin{table}[t]
\centering
\footnotesize
\caption{Results obtained on test datasets with and without division on frequencies as described in Section \ref{subsubsec:itte}}.
\label{tab:unified_processing}
\begin{tabular}{ c|| c c c}
  \hline
  \textit{\textbf{Dataset}} & Metric & Without Division & With Division\\
 \hline
 \multirow{2}{*}{Set5} & PSNR & 32.45 & \textbf{33.10}\\
 \cline{2-4}
 & SSIM & 0.8976 & \textbf{0.9092}\\
 \hline
 \multirow{2}{*}{Set14} & PSNR & 28.81 & \textbf{29.21}\\
 \cline{2-4}
 & SSIM & 0.7858 & \textbf{0.7929}\\
 \hline
 \multirow{2}{*}{B100} & PSNR & 27.73 & \textbf{27.98}\\
 \cline{2-4}
 & SSIM & 0.7417 & \textbf{0.7519}\\
 \hline 
 \multirow{2}{*}{Urban100} & PSNR & 26.66 & \textbf{27.51}\\
 \cline{2-4}
 & SSIM & 0.8035 & \textbf{0.8197}\\
 \hline 
 \multirow{2}{*}{Manga109} & PSNR & 31.02 & \textbf{32.17}\\
 \cline{2-4}
 & SSIM & 0.9146 & \textbf{0.9247}\\
 \hline 
 \end{tabular}
\end{table}

\subsubsection{Effectiveness of dividing the problem on frequency}
\label{subsubsec:itte}
To facilitate an effective division of the problem into sub-problems and to generate sharper upsampling results, WDN divides the problem of predicting all the frequencies of the upsampled image into the two problems of separately predicting the high-frequency and low-frequency details respectively. To verify the effectiveness of this division, we retrain and test WDN without separating the high-frequency and low-frequency details. Removal of the network components that participate in the prediction of the high-frequency details will result in the reduction of the number of trainable parameters. This will subsequently lead to an unfair comparison, and the numbers might not reflect the effect of frequency division. Hence, to make sure that the number of trainable parameters remains the same in this unified configuration, we first remove the frequency separator from WDN's architecture, as shown in Fig. \ref{fig:model}. Next, we replicate the input image and send it into the parallel networks that were originally designed for separately processing different frequencies. Ground-truths are also changed appropriately, wherever required. The results obtained with this configuration is shown in Table. \ref{tab:unified_processing}. It can be inferred from the table that separately and specifically modelling the high-frequency details is useful. 

\subsubsection{Effectiveness of dividing the problem on scale}
\label{subsubsec:edps}
To further facilitate the effective division of the given problem into sub-problems so as to make multiple sub-network (that can execute in parallel) learn to solve specific sub-problems with more expertise, WDN divides the problem of 4$\times$ upsampling into two successive problems of 2$\times$ upsampling. To verify the effectiveness of this division, we retrain and test WDN as a single 4$\times$ upsampling problem rather than two 2$\times$ upsampling problems. To conduct this experiment, we modify the architecture of WDN as follows. Space-to-depth operator (shown in Fig. \ref{fig:model}) is set to a block size of two. Due to this change, the space-to-depth operator generates four channels of high-frequency and four channels of low-frequency rather than 16, 16 channels. With only four channels, Stage-1 can be removed from WDN, and WDN can directly predict one 4$\times$ high-frequency channel and one 4$\times$ low-frequency channel with the help of Stage-2. All other configurations and Stage-3 remains the same as before. The results obtained with this change are shown in Table \ref{tab:division_scale}. 

It can be inferred from the table that the division of the problem into 11 sub-problems is indeed effective as opposed to division into three sub-problems (after removing eight sub-problems of Stage-1). The results further strengthen our claim that is dividing a problem into sub-problems and then solving them with multiple sub-networks, make the sub-networks gain more expertise in solving those sub-problems, eventually generating better sub-solutions that ultimately leads to better end solution.

\begin{table}[t]
\centering
\footnotesize
\caption{Results obtained on test datasets with and without division on the scale as described in Section \ref{subsubsec:edps}}.
\label{tab:division_scale}
\begin{tabular}{ c|| c c c}
  \hline
  \multirow{2}{*}{\textit{\textbf{Dataset}}} & \multirow{2}{*}{Metric} & No scale division & Two 2$\times$ division \\
  & & 3 sub-problems & 11 sub-problems\\
 \hline
 \multirow{2}{*}{Set5} & PSNR & 29.47 & \textbf{33.10}\\
 \cline{2-4}
 & SSIM & 0.8446 & \textbf{0.9092}\\
 \hline
 \multirow{2}{*}{Set14} & PSNR & 27.12 & \textbf{29.21}\\
 \cline{2-4}
 & SSIM & 0.7257 & \textbf{0.7929}\\
 \hline
 \multirow{2}{*}{B100} & PSNR & 26.49 & \textbf{27.98}\\
 \cline{2-4}
 & SSIM & 0.6993 & \textbf{0.7519}\\
 \hline 
 \multirow{2}{*}{Urban100} & PSNR & 24.93 & \textbf{27.51}\\
 \cline{2-4}
 & SSIM & 0.7553 & \textbf{0.8197}\\
 \hline 
 \multirow{2}{*}{Manga109} & PSNR & 27.77 & \textbf{32.17}\\
 \cline{2-4}
 & SSIM & 0.8689 & \textbf{0.9247}\\
 \hline 
 \end{tabular}
\end{table}
\begin{table}[t]
\centering
\footnotesize
\caption{Results obtained on test datasets with and without the attention blocks.}
\label{tab:attn}
\begin{tabular}{ c|| c c c}
  \hline
  \multirow{1}{*}{\textit{\textbf{Dataset}}} & \multirow{1}{*}{Metric} & Without Attention & With Attention\\

 \hline
 \multirow{2}{*}{Set5} & PSNR & 32.31 & \textbf{33.10}\\
 \cline{2-4}
 & SSIM & 0.8973 & \textbf{0.9092}\\
 \hline
 \multirow{2}{*}{Set14} & PSNR & 28.86 & \textbf{29.21}\\
 \cline{2-4}
 & SSIM & 0.7892 & \textbf{0.7929}\\
 \hline
 \multirow{2}{*}{B100} & PSNR & 27.79 & \textbf{27.98}\\
 \cline{2-4}
 & SSIM & 0.7409 & \textbf{0.7519}\\
 \hline 
 \multirow{2}{*}{Urban100} & PSNR & 26.59 & \textbf{27.51}\\
 \cline{2-4}
 & SSIM & 0.8014 & \textbf{0.8197}\\
 \hline 
 \multirow{2}{*}{Manga109} & PSNR & 31.29 & \textbf{32.17}\\
 \cline{2-4}
 & SSIM & 0.9153 & \textbf{0.9247}\\
 \hline 
 \end{tabular}
\end{table}

\subsubsection{Effectiveness of the attention mechanism}
\label{subsubsec:eotam}
WDN has extensively deployed shared attention blocks that evaluate the relative importance of pixels (located at corresponding locations in two or more feature-maps/channels). The computed relative importance help in weighing multiple sub-solutions to generate a single solution. To analyse the effectiveness of the attention blocks, we disable the attention mechanism and retrain the network. Table \ref{tab:attn} records the results on test datasets after making the said change.

It can be inferred from the table that the attention blocks are pivotal in WDN, and without these blocks, a drop in the performance is observed. A possible reason for this behaviour might be that the network finds it hard to implicitly learn the relative importance of pixels present in different sub-solutions while combining them to generate a single solution.

\subsubsection{Effectiveness of the divide-and-conquer based network design}
\label{subsubsec:eodac}
WDN is parametrically heavier network as compared to the state-of-the-art. To ascertain that the performance improvement obtained by WDN over the state-of-the-art is indeed due to the proposed divide-and-conquer based approach (i.e. division based on frequency and division based on scale) and not due to the parametric heaviness of WDN, we conduct an additional experiment. In this experiment, we replace the entire divide-and-conquer based network design of WDN with a much simpler linear design built with a series of (3$\times$3) Conv-2D layers followed by proposed pixel calibration. We add a Conv2D-Transpose layer with Sigmoid activation at the end to perform 4$\times$ upsampling. The network depth is set to 547 layers to make the number of parameters in the linear network, comparable to that of WDN. We observe that the linearly designed network does not converge, possibly due to the increased depth and use of Sigmoid in the pixel calibration. This experiment gives a clear indication that merely increasing the number of parameters of a network is ineffective to bring any performance improvement and that the parameters should be arranged in an effectively designed and intelligently crafted network such as WDN. We do mention that there can be other network designs that might have lesser parameters than WDN and that give comparable prediction quality. However, the design of such networks is beyond the scope of this work.

\subsubsection{Pixel calibration layer effectiveness}
\label{subsubsec:mhw}
For this analysis, we replace the proposed calibration layer with other typically used alternatives: 1) ReLU, 2) ReLU + Batch Norm., and 3) \textit{Srivastava \etal} \cite{NIPS2015_5850} version. The network has been retrained and retested after each change, and the results have been recorded in Table \ref{tab:pc_study}. The results indicate that the model significantly underperforms with only ReLU. Some improvement is observed with ReLU and Batch Normalisation. Much more improvement is observed with the version of \textit{Srivastava et al.}, and the highest results are obtained with our proposed layer as compared to other alternatives. These results reflect the effectiveness of the proposed calibration layer in the identification of the pixel relevance for super-resolution.

\subsubsection{High-frequency extractor choice: Sobel}
\label{subsec:sobel}
We use Sobel \cite{sobel19683x3} proposed in the work of Duda \etal \cite{duda2012pattern} for high-frequency extraction. It is a first-order derivative filter with $2_{(3\times3)}$ kernels (i.e. two 3$\times$3 kernels). It might not perform better than the higher-order alternatives (e.g. LoG \cite{marr1980theory}), but is computationally cheaper and has a lesser sensitivity to noise than higher-order filters. Comparing it with other first-order filters, Roberts \cite{roberts1963machine} with $2_{(2\times2)}$ kernels is computationally cheaper, but also performs lower than Sobel (as shown by Pratt \cite{10.5555/130597}). Prewitt \cite{prewitt1970object} with $2_{(3\times3)}$ kernels provide better results than Sobel, but only when the image is noiseless and well contrasted, thus reducing its genericity as compared to Sobel (ref. Adlakha \etal \cite{adlakha2016analytical}). The filters of Kirsch \cite{KIRSCH1971315}, Robinson \cite{ROBINSON1977492}, Frei-Chen \cite{1674733}, Nevatia-Babu \cite{NEVATIA1980257}, and Canny \cite{canny1983finding,4767851} might perform equally or slightly better than Sobel, but all of them are computationally expensive due to their $8_{(3\times3)}$, $8_{(3\times3)}$, $9_{(3\times3)}$, $12_{(5\times5)}$ kernels or algorithmic formulation, respectively (ref. Acharjya \etal \cite{acharjya2012study}). Based on these studies, we selected Sobel for WDN. It might not be the best but certainly is an optimal filter due to its performance and simplicity.

\begin{table}[t]
\footnotesize
\caption{Results after replacing the pixel calibration with: 1. ReLU, 2. ReLU + Batch normalisation, 3. \textit{Srivastava \etal} \cite{NIPS2015_5850}. 4. represents the proposed Pixel calibration.}
\label{tab:pc_study}
\centering
\begin{tabular}{ c|| c c c c c}
\hline
\textit{\textbf{Dataset}} & Config.$\rightarrow$ & 1 & 2 & 3 & 4\\
\hline
\multirow{2}{*}{Set5} & PSNR & 31.37 & 32.45 & 32.59 &  \textbf{33.10}\\
\cline{2-6}
& SSIM  & 0.8833 & 0.8984 & 0.9002 & \textbf{0.9092}\\
\hline
\multirow{2}{*}{Set14} & PSNR & 27.95 & 28.74 & 28.84 &  \textbf{29.21}\\
\cline{2-6}
& SSIM & 0.7667 & 0.7859 & 0.7877 & \textbf{0.7929}\\
\hline
\multirow{2}{*}{B100} & PSNR & 27.15 & 27.58 & 27.69 & \textbf{27.98}\\
\cline{2-6}
& SSIM  & 0.7246 & 0.7401 & 0.7414 &  \textbf{0.7519}\\
\hline 
\multirow{2}{*}{Urban100} & PSNR & 25.15 & 26.63 & 26.72  & \textbf{27.51}\\
\cline{2-6}
& SSIM  & 0.7532 & 0.8034 & 0.8054 & \textbf{0.8197}\\
\hline 
\multirow{2}{*}{Manga109} & PSNR & 28.79 & 30.83 & 30.91 & \textbf{32.17}\\
\cline{2-6}
& SSIM  & 0.8863 & 0.9131 & 0.9148 & \textbf{0.9247}\\
\hline
\end{tabular}
\end{table}

\section{Summary and future work}
\label{sec:conclude}
A wide and deep network (WDN) designed on the divide-and-conquer design paradigm has been proposed in this work. To solve the 4$\times$ image super-resolution problem, we divided the problem into three disjoint sets of 11 sub-problems, with each set having some simultaneously solvable subproblems. The division into sub-problems has been primarily made based on \textit{`Upsampling scale'} 4$\times$ upsampling gets divided into two 2$\times$ upsampling in sequence and \textit{`Frequency'} high-frequency, and low-frequency channels have been predicted separately. A wide and deep network abbreviated as WDN with pixel calibration layer has been designed to solve these sub-problems. We demonstrated that our approach towards solving the super-resolution problem gives better results (qualitatively and quantitatively) than state-of-the-arts on five publicly available datasets. Extensive ablation studies, empirically support the efficacy of all the components/ideas used in our work.

The idea of approaching a problem with divide-and-conquer along with a wide and deep network can be applied to other problems like video super-resolution, and deblurring, among others. Our future plan includes work to solve these problems.

\section*{Acknowledgement}
Authors would like to thank the JSTSP reviewers and editors for providing constructive feedback about the work. Authors would also like to thank the TFRC (TensorFlow Research Cloud) for giving free access to cloud Tensor Processing Units (TPUs).

\bibliographystyle{IEEEtran}
\bibliography{refs}

\begin{thebibliography}{10}
\providecommand{\url}[1]{#1}
\csname url@samestyle\endcsname
\providecommand{\newblock}{\relax}
\providecommand{\bibinfo}[2]{#2}
\providecommand{\BIBentrySTDinterwordspacing}{\spaceskip=0pt\relax}
\providecommand{\BIBentryALTinterwordstretchfactor}{4}
\providecommand{\BIBentryALTinterwordspacing}{\spaceskip=\fontdimen2\font plus
\BIBentryALTinterwordstretchfactor\fontdimen3\font minus
  \fontdimen4\font\relax}
\providecommand{\BIBforeignlanguage}[2]{{%
\expandafter\ifx\csname l@#1\endcsname\relax
\typeout{** WARNING: IEEEtran.bst: No hyphenation pattern has been}%
\typeout{** loaded for the language `#1'. Using the pattern for}%
\typeout{** the default language instead.}%
\else
\language=\csname l@#1\endcsname
\fi
#2}}
\providecommand{\BIBdecl}{\relax}
\BIBdecl

\bibitem{10.1007/978-3-030-01234-2_18}
Y.~{Zhang}, K.~{Li}, K.~{Li}, L.~{Wang}, B.~{Zhong}, and Y.~{Fu}, ``Image
  super-resolution using very deep residual channel attention networks,'' in
  \emph{ECCV}, V.~Ferrari, M.~Hebert, C.~Sminchisescu, and Y.~Weiss, Eds.,
  2018, pp. 294--310.

\bibitem{Li_2018_ECCV}
J.~Li, F.~Fang, K.~Mei, and G.~Zhang, ``Multi-scale residual network for image
  super-resolution,'' in \emph{ECCV}, V.~Ferrari, M.~Hebert, C.~Sminchisescu,
  and Y.~Weiss, Eds., 2018, pp. 527--542.

\bibitem{Dai_2019_CVPR}
T.~{Dai}, J.~{Cai}, Y.~{Zhang}, S.~{Xia}, and L.~{Zhang}, ``Second-order
  attention network for single image super-resolution,'' in \emph{CVPR}, 2019,
  pp. 11\,057--11\,066.

\bibitem{5459271}
D.~{Glasner}, S.~{Bagon}, and M.~{Irani}, ``Super-resolution from a single
  image,'' in \emph{ICCV}, Sep. 2009, pp. 349--356.

\bibitem{5466111}
J.~{Yang}, J.~{Wright}, T.~S. {Huang}, and Y.~{Ma}, ``Image super-resolution
  via sparse representation,'' \emph{IEEE TIP}, vol.~19, no.~11, pp.
  2861--2873, Nov 2010.

\bibitem{6751349}
R.~{Timofte}, V.~{De}, and L.~V. {Gool}, ``Anchored neighborhood regression for
  fast example-based super-resolution,'' in \emph{ICCV}, Dec 2013, pp.
  1920--1927.

\bibitem{7299003}
S.~{Schulter}, C.~{Leistner}, and H.~{Bischof}, ``Fast and accurate image
  upscaling with super-resolution forests,'' in \emph{CVPR}, June 2015, pp.
  3791--3799.

\bibitem{7780576}
W.~{Shi}, J.~{Caballero}, F.~{Huszár}, J.~{Totz}, A.~P. {Aitken}, R.~{Bishop},
  D.~{Rueckert}, and Z.~{Wang}, ``Real-time single image and video
  super-resolution using an efficient sub-pixel convolutional neural network,''
  in \emph{CVPR}, June 2016, pp. 1874--1883.

\bibitem{10.1007/978-3-319-46475-6_43}
J.~Johnson, A.~Alahi, and L.~Fei-Fei, ``Perceptual losses for real-time style
  transfer and super-resolution,'' in \emph{ECCV}, B.~Leibe, J.~Matas, N.~Sebe,
  and M.~Welling, Eds., 2016, pp. 694--711.

\bibitem{7780551}
J.~{Kim}, J.~K. {Lee}, and K.~M. {Lee}, ``Accurate image super-resolution using
  very deep convolutional networks,'' in \emph{CVPR}, June 2016, pp.
  1646--1654.

\bibitem{10.1007/978-3-319-46454-1_20}
X.~Yu and F.~Porikli, ``Ultra-resolving face images by discriminative
  generative networks,'' in \emph{ECCV}, B.~Leibe, J.~Matas, N.~Sebe, and
  M.~Welling, Eds., 2016, pp. 318--333.

\bibitem{7780550}
J.~{Kim}, J.~K. {Lee}, and K.~M. {Lee}, ``Deeply-recursive convolutional
  network for image super-resolution,'' in \emph{CVPR}, June 2016, pp.
  1637--1645.

\bibitem{8099781}
Y.~{Tai}, J.~{Yang}, and X.~{Liu}, ``Image super-resolution via deep recursive
  residual network,'' in \emph{CVPR}, July 2017, pp. 2790--2798.

\bibitem{8100101}
W.~{Lai}, J.~{Huang}, N.~{Ahuja}, and M.~{Yang}, ``Deep laplacian pyramid
  networks for fast and accurate super-resolution,'' in \emph{CVPR}, July 2017,
  pp. 5835--5843.

\bibitem{8578422}
J.~{Pan}, S.~{Liu}, D.~{Sun}, J.~{Zhang}, Y.~{Liu}, J.~{Ren}, Z.~{Li},
  J.~{Tang}, H.~{Lu}, Y.~{Tai}, and M.~{Yang}, ``Learning dual convolutional
  neural networks for low-level vision,'' in \emph{CVPR}, June 2018, pp.
  3070--3079.

\bibitem{10.1007/978-3-030-01249-6_16}
N.~Ahn, B.~Kang, and K.-A. Sohn, ``Fast, accurate, and lightweight
  super-resolution with cascading residual network,'' in \emph{ECCV},
  V.~Ferrari, M.~Hebert, C.~Sminchisescu, and Y.~Weiss, Eds., 2018, pp.
  256--272.

\bibitem{8237743}
M.~S.~M. {Sajjadi}, B.~{Schölkopf}, and M.~{Hirsch}, ``Enhancenet: Single
  image super-resolution through automated texture synthesis,'' in \emph{ICCV},
  Oct 2017, pp. 4501--4510.

\bibitem{8099502}
C.~{Ledig}, L.~{Theis}, F.~{Huszár}, J.~{Caballero}, A.~{Cunningham},
  A.~{Acosta}, A.~{Aitken}, A.~{Tejani}, J.~{Totz}, Z.~{Wang}, and W.~{Shi},
  ``Photo-realistic single image super-resolution using a generative
  adversarial network,'' in \emph{CVPR}, July 2017, pp. 105--114.

\bibitem{8237298}
X.~{Xu}, D.~{Sun}, J.~{Pan}, Y.~{Zhang}, H.~{Pfister}, and M.~{Yang},
  ``Learning to super-resolve blurry face and text images,'' in \emph{ICCV},
  Oct 2017, pp. 251--260.

\bibitem{8237843}
R.~{Dahl}, M.~{Norouzi}, and J.~{Shlens}, ``Pixel recursive super resolution,''
  in \emph{ICCV}, Oct 2017, pp. 5449--5458.

\bibitem{8237748}
Y.~{Tai}, J.~{Yang}, X.~{Liu}, and C.~{Xu}, ``Memnet: A persistent memory
  network for image restoration,'' in \emph{ICCV}, Oct 2017, pp. 4549--4557.

\bibitem{8014885}
B.~{Lim}, S.~{Son}, H.~{Kim}, S.~{Nah}, and K.~M. {Lee}, ``Enhanced deep
  residual networks for single image super-resolution,'' in \emph{CVPRW}, July
  2017, pp. 1132--1140.

\bibitem{10.1007/978-3-030-01231-1_6}
H.~Zheng, M.~Ji, H.~Wang, Y.~Liu, and L.~Fang, ``Crossnet: An end-to-end
  reference-based super resolution network using cross-scale warping,'' in
  \emph{ECCV}, V.~Ferrari, M.~Hebert, C.~Sminchisescu, and Y.~Weiss, Eds.,
  2018, pp. 87--104.

\bibitem{8578180}
Z.~{Hui}, X.~{Wang}, and X.~{Gao}, ``Fast and accurate single image
  super-resolution via information distillation network,'' in \emph{CVPR}, June
  2018, pp. 723--731.

\bibitem{10.1007/978-3-030-01237-3_32}
J.~Li, F.~Fang, K.~Mei, and G.~Zhang, ``Multi-scale residual network for image
  super-resolution,'' in \emph{ECCV}, V.~Ferrari, M.~Hebert, C.~Sminchisescu,
  and Y.~Weiss, Eds., 2018, pp. 527--542.

\bibitem{Zhong:2018:JSL:3326943.3326959}
Z.~Zhong, T.~Shen, Y.~Yang, C.~Zhang, and Z.~Lin, ``Joint sub-bands learning
  with clique structures for wavelet domain super-resolution,'' in \emph{NIPS},
  2018, pp. 165--175.

\bibitem{10.1007/978-3-030-01231-1_12}
A.~Bulat, J.~Yang, and G.~Tzimiropoulos, ``To learn image super-resolution, use
  a gan to learn how to do image degradation first,'' in \emph{ECCV},
  V.~Ferrari, M.~Hebert, C.~Sminchisescu, and Y.~Weiss, Eds., 2018, pp.
  187--202.

\bibitem{8578442}
K.~{Zhang}, W.~{Zuo}, and L.~{Zhang}, ``Learning a single convolutional
  super-resolution network for multiple degradations,'' in \emph{CVPR}, June
  2018, pp. 3262--3271.

\bibitem{ROMANIUK19931105}
S.~G. Romaniuk and L.~O. Hall, ``Divide and conquer neural networks,''
  \emph{Neural Networks}, vol.~6, no.~8, pp. 1105 -- 1116, 1993.

\bibitem{nowak2018divide}
\BIBentryALTinterwordspacing
A.~Nowak, D.~Folqué, and J.~Bruna, ``Divide and conquer networks,'' in
  \emph{International Conference on Learning Representations}, 2018. [Online].
  Available: \url{https://openreview.net/forum?id=B1jscMbAW}
\BIBentrySTDinterwordspacing

\bibitem{ghosh2018divideandconquer}
D.~Ghosh, A.~Singh, A.~Rajeswaran, V.~Kumar, and S.~Levine,
  ``Divide-and-conquer reinforcement learning,'' in \emph{International
  Conference on Learning Representations}, 2018.

\bibitem{9010947}
C.~H. {Lin}, C.~{Chang}, Y.~{Chen}, D.~{Juan}, W.~{Wei}, and H.~{Chen},
  ``Coco-gan: Generation by parts via conditional coordinating,'' in
  \emph{ICCV}, 2019, pp. 4511--4520.

\bibitem{AAAI2020}
S.~Y. Kim, J.~Oh, and M.~Kim, ``Jsi-gan: Gan-based joint super-resolution and
  inverse tone-mapping with pixel-wise task-specific filters for uhd hdr
  video,'' in \emph{AAAI}, 2020, pp. 11\,287--11\,295.

\bibitem{huang2019divideandconquer}
Z.~Huang, D.~P. Paudel, G.~Li, J.~Wu, R.~Timofte, and L.~V. Gool,
  ``Divide-and-conquer adversarial learning for high-resolution image and video
  enhancement,'' 2019.

\bibitem{8578276}
W.~{Han}, S.~{Chang}, D.~{Liu}, M.~{Yu}, M.~{Witbrock}, and T.~S. {Huang},
  ``Image super-resolution via dual-state recurrent networks,'' in \emph{CVPR},
  June 2018, pp. 1654--1663.

\bibitem{Gu_2019_CVPR}
J.~{Gu}, H.~{Lu}, W.~{Zuo}, and C.~{Dong}, ``Blind super-resolution with
  iterative kernel correction,'' in \emph{CVPR}, 2019, pp. 1604--1613.

\bibitem{Zhou_2019_ICCV1}
R.~{Zhou} and S.~{Süsstrunk}, ``Kernel modeling super-resolution on real
  low-resolution images,'' in \emph{ICCV)}, 2019, pp. 2433--2443.

\bibitem{Zhang_2019_CVPR1}
K.~{Zhang}, W.~{Zuo}, and L.~{Zhang}, ``Deep plug-and-play super-resolution for
  arbitrary blur kernels,'' in \emph{CVPR}, 2019, pp. 1671--1681.

\bibitem{Qiu_2019_ICCV}
Y.~{Qiu}, R.~{Wang}, D.~{Tao}, and J.~{Cheng}, ``Embedded block residual
  network: A recursive restoration model for single-image super-resolution,''
  in \emph{ICCV}, 2019, pp. 4179--4188.

\bibitem{He_2019_CVPR}
X.~{He}, Z.~{Mo}, P.~{Wang}, Y.~{Liu}, M.~{Yang}, and J.~{Cheng},
  ``Ode-inspired network design for single image super-resolution,'' in
  \emph{CVPR}, 2019, pp. 1732--1741.

\bibitem{8578277}
M.~{Haris}, G.~{Shakhnarovich}, and N.~{Ukita}, ``Deep back-projection networks
  for super-resolution,'' in \emph{CVPR}, June 2018, pp. 1664--1673.

\bibitem{Hu_2019_CVPR}
X.~{Hu}, H.~{Mu}, X.~{Zhang}, Z.~{Wang}, T.~{Tan}, and J.~{Sun}, ``Meta-sr: A
  magnification-arbitrary network for super-resolution,'' in \emph{CVPR}, 2019,
  pp. 1575--1584.

\bibitem{Li_2019_CVPR}
Z.~{Li}, J.~{Yang}, Z.~{Liu}, X.~{Yang}, G.~{Jeon}, and W.~{Wu}, ``Feedback
  network for image super-resolution,'' in \emph{CVPR}, 2019, pp. 3862--3871.

\bibitem{8578427}
A.~{Shocher}, N.~{Cohen}, and M.~{Irani}, ``Zero-shot super-resolution using
  deep internal learning,'' in \emph{CVPR}, June 2018, pp. 3118--3126.

\bibitem{10.1007/978-3-030-01270-0_27}
S.-J. Park, H.~Son, S.~Cho, K.-S. Hong, and S.~Lee, ``Srfeat: Single image
  super-resolution with feature discrimination,'' in \emph{ECCV}, V.~Ferrari,
  M.~Hebert, C.~Sminchisescu, and Y.~Weiss, Eds., 2018, pp. 455--471.

\bibitem{Zhang_2019_ICCV}
W.~{Zhang}, Y.~{Liu}, C.~{Dong}, and Y.~{Qiao}, ``Ranksrgan: Generative
  adversarial networks with ranker for image super-resolution,'' in
  \emph{ICCV}, 2019, pp. 3096--3105.

\bibitem{Rad_2019_ICCV}
M.~S. {Rad}, B.~{Bozorgtabar}, U.~{Marti}, M.~{Basler}, H.~K. {Ekenel}, and
  J.~{Thiran}, ``Srobb: Targeted perceptual loss for single image
  super-resolution,'' in \emph{ICCV}, 2019, pp. 2710--2719.

\bibitem{9053890}
H.~{Liu}, Z.~{Lu}, W.~{Shi}, and J.~{Tu}, ``A fast and accurate
  super-resolution network using progressive residual learning,'' in
  \emph{ICASSP}, 2020, pp. 1818--1822.

\bibitem{9054003}
K.~{Liu}, Z.~{Han}, J.~{Chen}, C.~{Liu}, J.~{Chen}, and Z.~{Wang}, ``Image
  super-resolution using residual global context network,'' in \emph{ICASSP},
  2020, pp. 2633--2637.

\bibitem{9054019}
Z.~{Hou} and S.~{Kung}, ``Efficient image super resolution via channel
  discriminative deep neural network pruning,'' in \emph{ICASSP}, 2020, pp.
  3647--3651.

\bibitem{KIM2020}
J.-H. Kim, J.-H. Choi, M.~Cheon, and J.-S. Lee, ``Mamnet: Multi-path adaptive
  modulation network for image super-resolution,'' \emph{Neurocomputing}, 2020.

\bibitem{9053516}
M.~{Zou}, J.~{Tang}, and G.~{Wu}, ``Low complexity single image
  super-resolution with channel splitting and fusion network,'' in
  \emph{ICASSP}, 2020, pp. 2398--2402.

\bibitem{9066958}
W.~{Wang}, G.~{Wu}, W.~{Cai}, L.~{Zeng}, and J.~{Chen}, ``Robust prior-based
  single image super resolution under multiple gaussian degradations,''
  \emph{IEEE Access}, vol.~8, pp. 74\,195--74\,204, 2020.

\bibitem{Qin2020}
D.~Qin and X.~Gu, ``Single-image super-resolution with multilevel residual
  attention network,'' \emph{Neural Computing and Applications}, Apr 2020.

\bibitem{Wu2020}
Q.~Wu, C.~Fan, Y.~Li, Y.~Li, and J.~Hu, ``A novel perceptual loss function for
  single image super-resolution,'' \emph{Multimedia Tools and Applications},
  May 2020.

\bibitem{IRANI1993324}
M.~Irani and S.~Peleg, ``Motion analysis for image enhancement: Resolution,
  occlusion, and transparency,'' \emph{Journal of Visual Communication and
  Image Representation}, vol.~4, no.~4, pp. 324 -- 335, 1993.

\bibitem{1284395}
{Zhou Wang}, A.~C. {Bovik}, H.~R. {Sheikh}, and E.~P. {Simoncelli}, ``Image
  quality assessment: from error visibility to structural similarity,''
  \emph{IEEE TIP}, vol.~13, no.~4, pp. 600--612, April 2004.

\bibitem{space-to-depth}
{Space-to-depth}, ``{TensorFlow},''
  \url{https://www.tensorflow.org/api_docs/python/tf/nn/space_to_depth},
  accessed: 2020-05-15.

\bibitem{7444187}
A.~{Kappeler}, S.~{Yoo}, Q.~{Dai}, and A.~K. {Katsaggelos}, ``Video
  super-resolution with convolutional neural networks,'' \emph{IEEE TCI},
  vol.~2, no.~2, pp. 109--122, June 2016.

\bibitem{8237536}
D.~{Liu}, Z.~{Wang}, Y.~{Fan}, X.~{Liu}, Z.~{Wang}, S.~{Chang}, and T.~{Huang},
  ``Robust video super-resolution with learned temporal dynamics,'' in
  \emph{ICCV}, Oct 2017, pp. 2526--2534.

\bibitem{NIPS2015_5850}
R.~K. Srivastava, K.~Greff, and J.~Schmidhuber, ``Training very deep
  networks,'' in \emph{NIPS}, 2015, p. 2377–2385.

\bibitem{8578360}
Y.~{Zhang}, Y.~{Tian}, Y.~{Kong}, B.~{Zhong}, and Y.~{Fu}, ``Residual dense
  network for image super-resolution,'' in \emph{CVPR}, June 2018, pp.
  2472--2481.

\bibitem{pmlr-v9-glorot10a}
X.~Glorot and Y.~Bengio, ``Understanding the difficulty of training deep
  feedforward neural networks,'' in \emph{Proceedings of the Thirteenth
  International Conference on Artificial Intelligence and Statistics}, ser.
  Proceedings of Machine Learning Research, Y.~W. Teh and M.~Titterington,
  Eds., vol.~9.\hskip 1em plus 0.5em minus 0.4em\relax Chia Laguna Resort,
  Sardinia, Italy: PMLR, 13--15 May 2010, pp. 249--256.

\bibitem{DBLP:journals/corr/KingmaB14}
D.~P. Kingma and J.~Ba, ``Adam: {A} method for stochastic optimization,'' in
  \emph{ICLR}, 2015, pp.~--.

\bibitem{8014883}
R.~{Timofte}, E.~{Agustsson}, L.~V. {Gool} \emph{et~al.}, ``Ntire 2017
  challenge on single image super-resolution: Methods and results,'' in
  \emph{CVPRW}, July 2017, pp. 1110--1121.

\bibitem{set5}
M.~Bevilacqua, A.~Roumy, C.~Guillemot, and M.~line Alberi~Morel,
  ``Low-complexity single-image super-resolution based on nonnegative neighbor
  embedding,'' in \emph{BMVC}, 2012, pp. 135.1--135.10.

\bibitem{set14}
R.~Zeyde, M.~Elad, and M.~Protter, ``On single image scale-up using
  sparse-representations,'' in \emph{Curves and Surfaces}, J.-D. Boissonnat,
  P.~Chenin, A.~Cohen, C.~Gout, T.~Lyche, M.-L. Mazure, and L.~Schumaker,
  Eds.\hskip 1em plus 0.5em minus 0.4em\relax Springer Berlin Heidelberg, 2010,
  pp. 711--730.

\bibitem{937655}
D.~{Martin}, C.~{Fowlkes}, D.~{Tal}, and J.~{Malik}, ``A database of human
  segmented natural images and its application to evaluating segmentation
  algorithms and measuring ecological statistics,'' in \emph{ICCV}, vol.~2,
  July 2001, pp. 416--423 vol.2.

\bibitem{10.1007/978-3-319-16817-3_8}
R.~Timofte, V.~De Smet, and L.~Van Gool, ``A+: Adjusted anchored neighborhood
  regression for fast super-resolution,'' in \emph{ACCV}, D.~Cremers, I.~Reid,
  H.~Saito, and M.-H. Yang, Eds., 2015, pp. 111--126.

\bibitem{7299156}
J.~{Huang}, A.~{Singh}, and N.~{Ahuja}, ``Single image super-resolution from
  transformed self-exemplars,'' in \emph{CVPR}, June 2015, pp. 5197--5206.

\bibitem{Matsui2017}
Y.~Matsui, K.~Ito, Y.~Aramaki, A.~Fujimoto, T.~Ogawa, T.~Yamasaki, and
  K.~Aizawa, ``Sketch-based manga retrieval using manga109 dataset,''
  \emph{Multimedia Tools and Applications}, vol.~76, no.~20, pp.
  21\,811--21\,838, Oct 2017.

\bibitem{sobel19683x3}
I.~Sobel and G.~Feldman, ``A 3x3 isotropic gradient operator for image
  processing,'' \emph{a talk at the Stanford Artificial Project in}, pp.
  271--272, 1968.

\bibitem{duda2012pattern}
R.~O. Duda and P.~E. Hart, \emph{Pattern classification and Scene
  Analysis}.\hskip 1em plus 0.5em minus 0.4em\relax John Wiley \& Sons, 1973.

\bibitem{marr1980theory}
D.~Marr and E.~Hildreth, ``Theory of edge detection,'' \emph{Proceedings of the
  Royal Society of London. Series B. Biological Sciences}, vol. 207, no. 1167,
  pp. 187--217, 1980.

\bibitem{roberts1963machine}
L.~G. Roberts, ``Machine perception of three-dimensional solids,'' Ph.D.
  dissertation, Massachusetts Institute of Technology, 1963.

\bibitem{10.5555/130597}
W.~K. Pratt, \emph{Digital Image Processing (2nd Ed.)}.\hskip 1em plus 0.5em
  minus 0.4em\relax USA: John Wiley \& Sons, Inc., 1991.

\bibitem{prewitt1970object}
J.~M. Prewitt, ``Object enhancement and extraction,'' \emph{Picture processing
  and Psychopictorics}, vol.~10, no.~1, pp. 15--19, 1970.

\bibitem{adlakha2016analytical}
D.~Adlakha, D.~Adlakha, and R.~Tanwar, ``Analytical comparison between sobel
  and prewitt edge detection techniques,'' \emph{International Journal of
  Scientific \& Engineering Research}, vol.~7, no.~1, p.~4, 2016.

\bibitem{KIRSCH1971315}
R.~A. Kirsch, ``Computer determination of the constituent structure of
  biological images,'' \emph{Computers and Biomedical Research}, vol.~4, no.~3,
  pp. 315 -- 328, 1971.

\bibitem{ROBINSON1977492}
G.~S. Robinson, ``Edge detection by compass gradient masks,'' \emph{Computer
  Graphics and Image Processing}, vol.~6, no.~5, pp. 492 -- 501, 1977.

\bibitem{1674733}
{Frei} and {Chung-Ching Chen}, ``Fast boundary detection: A generalization and
  a new algorithm,'' \emph{IEEE Transactions on Computers}, vol. C-26, no.~10,
  pp. 988--998, 1977.

\bibitem{NEVATIA1980257}
R.~Nevatia and K.~R. Babu], ``Linear feature extraction and description,''
  \emph{Computer Graphics and Image Processing}, vol.~13, no.~3, pp. 257 --
  269, 1980.

\bibitem{canny1983finding}
J.~F. Canny, ``Finding edges and lines in images.'' MIT Cambridge AI Lab, Tech.
  Rep., 1983.

\bibitem{4767851}
J.~{Canny}, ``A computational approach to edge detection,'' \emph{IEEE TPAMI},
  vol. PAMI-8, no.~6, pp. 679--698, 1986.

\bibitem{acharjya2012study}
P.~P. Acharjya, R.~Das, and D.~Ghoshal, ``Study and comparison of different
  edge detectors for image segmentation,'' \emph{Global Journal of Computer
  Science and Technology}, 2012.

\end{thebibliography}
\newpage
\begin{IEEEbiography}[{\includegraphics[width=1in,height=1.25in,clip,keepaspectratio]{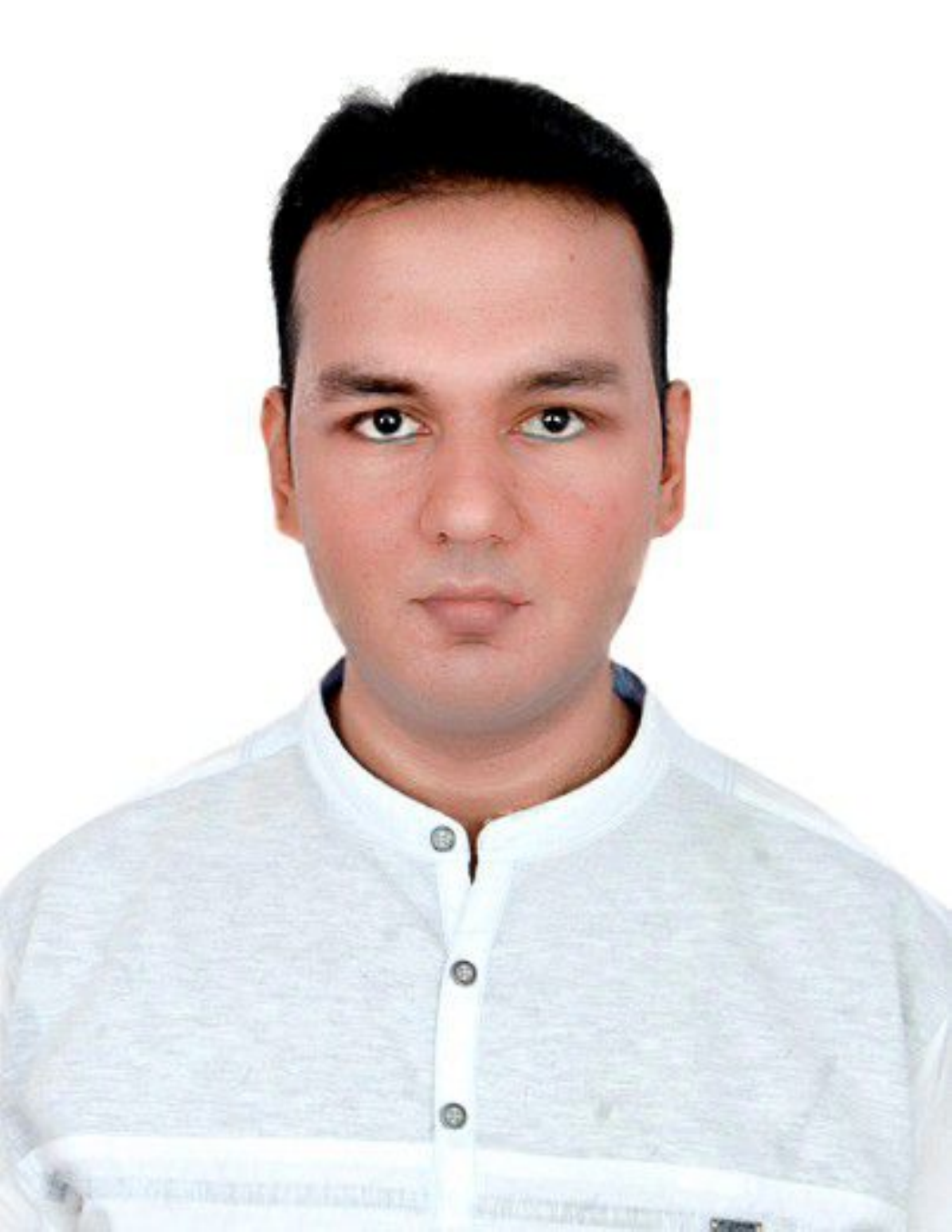}}]{Vikram Singh}
is a research scholar (PhD) under the supervision of Prof. Anurag Mittal in Computer Vision Lab, Department of Computer Science and Engineering, IIT-Madras. He completed his Masters in Technology from IIT-Patna in 2015 and Masters in Computer Application from Panjab University, Chandigarh in 2012. His current research interests include deep learning for Computer Vision, particularly restoration techniques such as super-resolution and deblurring, among others.
\end{IEEEbiography}
\vskip -30pt plus -1fil
\begin{IEEEbiography}[{\includegraphics[width=1in,height=1.25in,clip,keepaspectratio]{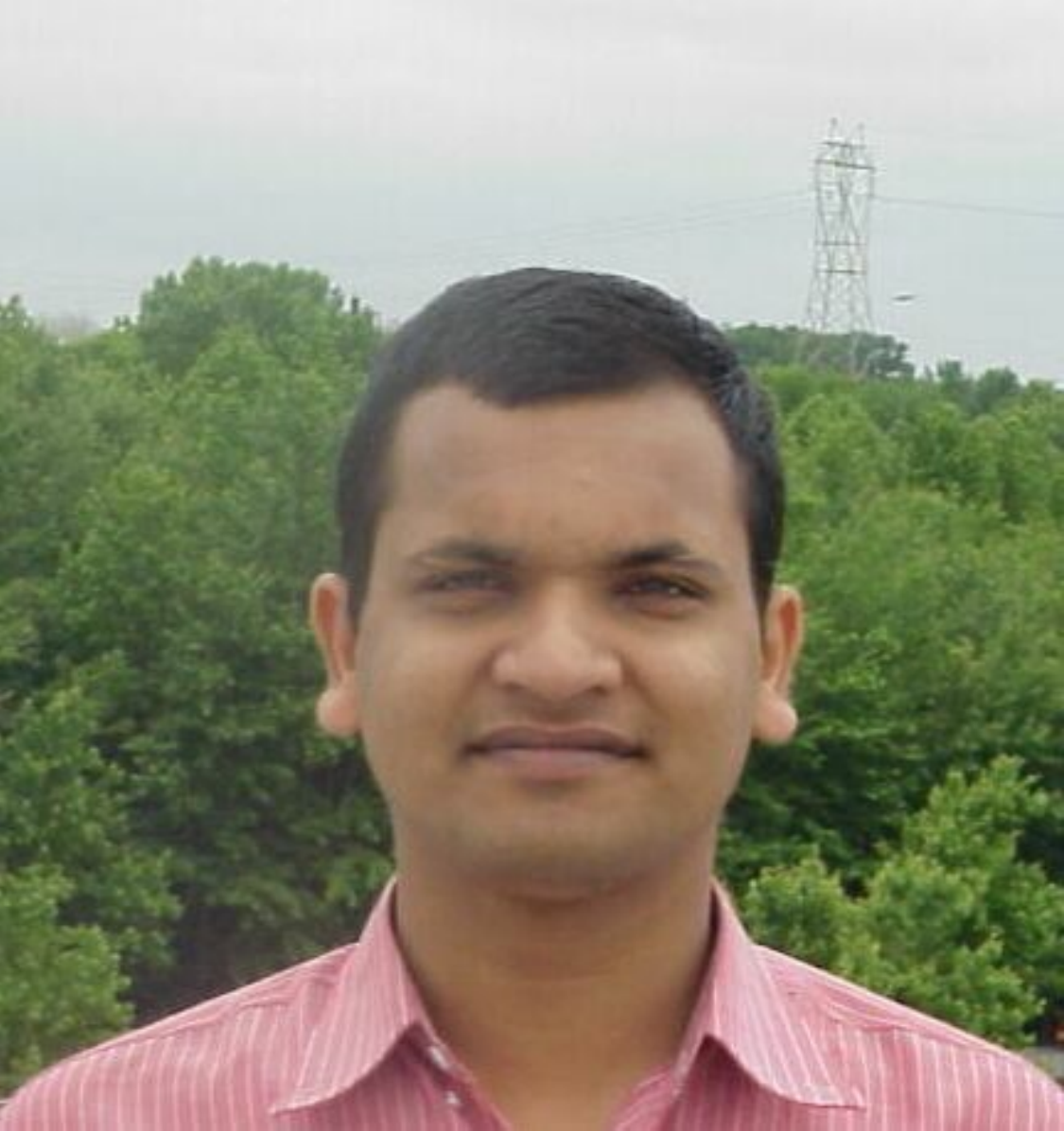}}]{Anurag Mittal}
is a professor in the CSE dept. at IIT-Madras where he heads the Computer Vision lab. He holds a PhD from the University of Maryland, College Park and an MS from Cornell University. His research interests are in all areas of Computer Vision, with a special interest in combining geometric and statistical inference techniques to build robust Computer Vision systems. He has published extensively in top venues of the field and has over 4000 citations. He is an Area Editor for CVIU and has been an Area Chair for ACCV and CVPR.
\end{IEEEbiography}

\end{document}